%% file: main.tex
\documentclass[11pt, a4paper, logo, copyright, nonumbering]{deepseek}
\usepackage[authoryear, sort&compress, round]{natbib}
\usepackage{dblfloatfix}
\usepackage{ulem}
\usepackage{caption}
\usepackage{dramatist}
\usepackage{xspace}
\usepackage{pifont}
\usepackage{multirow}
\usepackage{tcolorbox}
\usepackage{xltabular}
\usepackage{longtable}
\usepackage{hyperref}
\usepackage{amsfonts}
\usepackage{amsmath}
\usepackage{amssymb}
\usepackage{lineno}
\usepackage{algorithm}
\usepackage{algpseudocode}
\usepackage[bottom]{footmisc}
\usepackage{CJKutf8}
\usepackage{subfigure}
\usepackage{setspace}

\makeatletter
\def\@BTrule[#1]{%
  \ifx\longtable\undefined
    \let\@BTswitch\@BTnormal
  \else\ifx\hline\LT@hline
    \nobreak
    \let\@BTswitch\@BLTrule
  \else
     \let\@BTswitch\@BTnormal
  \fi\fi
  \global\@thisrulewidth=#1\relax
  \ifnum\@thisruleclass=\tw@\vskip\@aboverulesep\else
  \ifnum\@lastruleclass=\z@\vskip\@aboverulesep\else
  \ifnum\@lastruleclass=\@ne\vskip\doublerulesep\fi\fi\fi
  \@BTswitch}
\makeatother

\addto\extrasenglish{
}

 {\begin{list}{}%
         {\setlength{\leftmargin}{#1}}%
         \item[]%
 }
 {\end{list}}
 
\reportnumber{001} 

\renewcommand{\today}{}

\title{\centering DeepSeek-Coder: When the Large Language Model Meets Programming - The Rise of Code Intelligence}

\author[*]{
\small
\hspace{2.4em}
Daya Guo*$^{1}$,
Qihao Zhu$^{*1,2}$,
Dejian Yang$^{1}$,
Zhenda Xie$^{1}$, 
Kai Dong$^{1}$,
Wentao Zhang$^{1}$
\newline
Guanting Chen$^{1}$,
Xiao Bi $^{1}$,
Y. Wu$^{1}$,
Y.K. Li$^{1}$,
Fuli Luo$^{1}$,
Yingfei Xiong$^{2}$,
Wenfeng Liang$^{1}$ 
\\
\small
$^1$DeepSeek-AI \\
\small
$^2$Key Lab of HCST (PKU), MOE; SCS, Peking University \\
\small
\texttt{\{zhuqh, guodaya\}@deepseek.com} \\
\small
\url{https://github.com/deepseek-ai/DeepSeek-Coder}
}
\correspondingauthor{Core contributors, ordered alphabetically by the name.
}

\input{commands.tex}

\newcommand{\dscoder}{DeepSeek-Coder\xspace}
\newcommand{\dsbase}{DeepSeek-Coder-Base\xspace}
\newcommand{\dsins}{DeepSeek-Coder-Instruct\xspace}

\begin{abstract}

The rapid development of large language models has revolutionized code intelligence in software development. However, the predominance of closed-source models has restricted extensive research and development. To address this, we introduce the \dscoder series, a range of open-source code models with sizes from 1.3B to 33B, trained from scratch on 2 trillion tokens. These models are pre-trained on a high-quality project-level code corpus and employ a fill-in-the-blank task with a 16K window to enhance code generation and infilling. Our extensive evaluations demonstrate that \dscoder not only achieves state-of-the-art performance among open-source code models across multiple benchmarks but also surpasses existing closed-source models like Codex and GPT-3.5. Furthermore, \dscoder models are under a permissive license that allows for both research and unrestricted commercial use.

\end{abstract}

\begin{document}

\begin{CJK*}{UTF8}{gbsn}

\maketitle



\begin{figure}[h]
    \centering
    \includegraphics[width=0.99\linewidth]{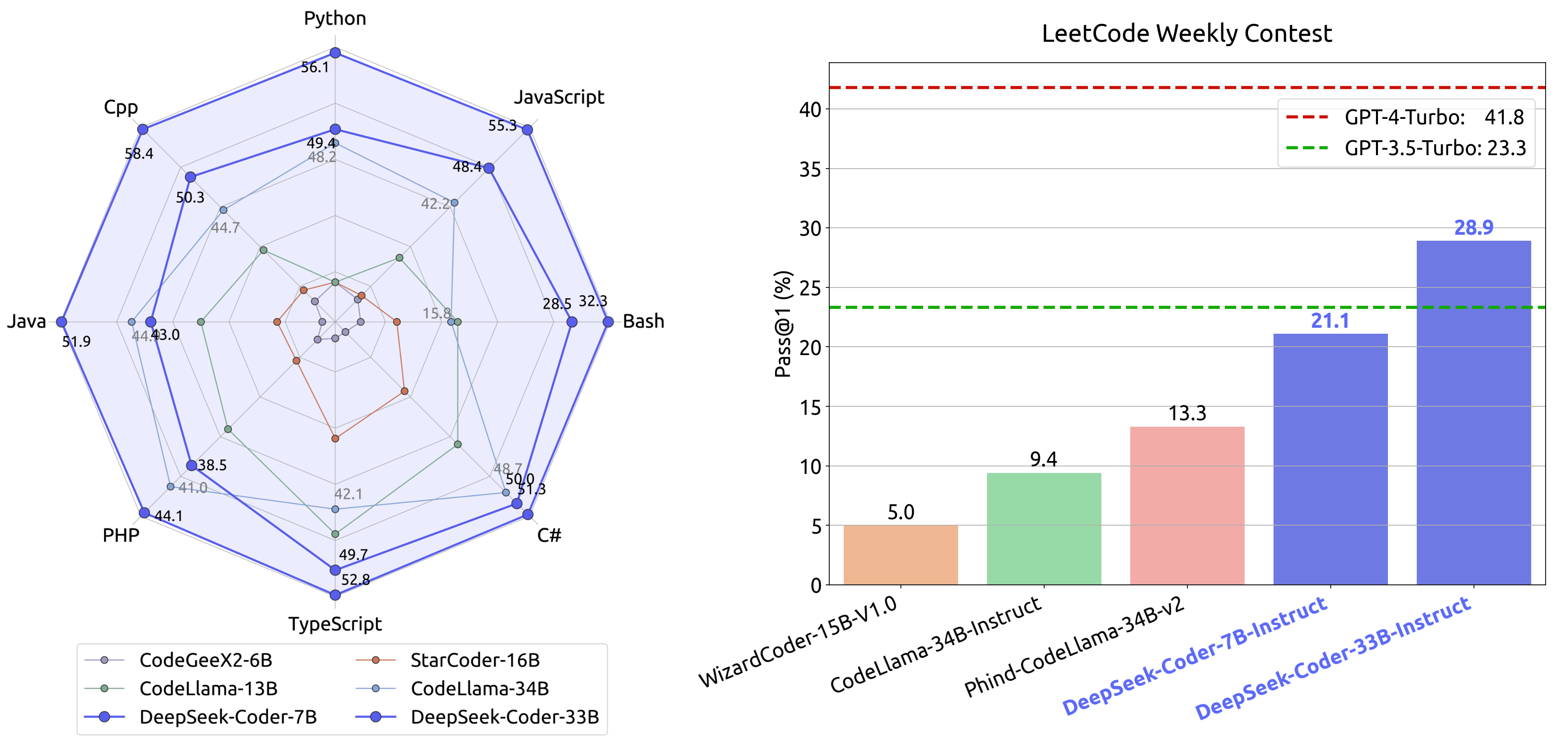}
    \caption{\centering The Performance of \dscoder}
\label{fig:data}
\end{figure}

\section{Introduction}
\label{par:intro}
The field of software development has been significantly transformed by the swift advancement of large language models ~\citep{touvron2023llama,openai2023gpt4}, which have brought about a new era of code intelligence. These models have the potential to automate and streamline many aspects of coding, from bug detection to code generation, thereby enhancing productivity and reducing the likelihood of human error. However, a major challenge in this field is the performance gap between open-source models ~\citep{roziere2023code,li2023starcoder,nijkamp2022codegen,wang2021codet5}  and closed-source models ~\citep{openai2023gpt4,gemini}. The giant closed-source models, while powerful, are often inaccessible to many researchers and developers due to their proprietary nature. 

In response to this challenge, we present the \dscoder series. This series comprises a range of open-source code models, varying in size from 1.3B to 33B, including the base version and instructed version for each size. Each model in the series has been trained from scratch on  2 trillion tokens sourced from 87 programming languages, ensuring a comprehensive understanding of coding languages and syntax. Besides, we attempt to organize the pre-training data at the repository level to enhance the pre-trained model's understanding capability within the context of cross-files within a repository. In addition to employing the next token prediction loss during pre-training, we have also incorporated the Fill-In-Middle (FIM) approach ~\citep{li2023starcoder,bavarian2022efficient}. This approach is designed to further bolster the model's code completion capabilities.
To meet the requirements of handling longer code inputs, we have extended the context length to 16K. This adjustment allows our models to handle more complex and extensive coding tasks, thereby increasing their versatility and applicability in various coding scenarios.

We have carried out comprehensive experiments using a variety of public code-related benchmarks. The findings reveal that among open-source models, \dsbase 33B consistently delivers superior performance across all benchmarks. Furthermore, \dsins 33B surpasses  \textit{OpenAI GPT-3.5 Turbo}  in the majority of the evaluation benchmarks, significantly narrowing the performance gap between \textit{OpenAI GPT-4}  and open-source models. Remarkably, despite having fewer parameters, \dsbase 7B demonstrates competitive performance when compared to models that are five times larger, such as CodeLlama-33B \citep{roziere2023code}. To summarize, our main contributions are:

\begin{itemize}[leftmargin=20pt]
    \item We introduce \dsbase and \dsins, our advanced code-focused large language models (LLMs). Developed through extensive training on an expansive code corpus, these models exhibit proficiency in understanding 87 programming languages. Additionally, they are available in various model scales to cater to a wide range of computational and application needs.
    \item  We make the first attempt to incorporate repository-level data construction during the pre-training phase of our models. We find that
    it can significantly boost the capability of cross-file code generation.  
    \item Our analysis rigorously examines the impact of FIM training strategies on the pretraining phase of code models. The outcomes of these comprehensive studies shed light on intriguing aspects of FIM configurations, offering valuable insights that significantly contribute to the enhancement and development of code pretrained models.
    \item We conduct extensive evaluations of our code LLMs against a wide array of benchmarks encompassing numerous code-related tasks. The findings demonstrate that \dsbase surpasses all existing open-source code LLMs across these benchmarks. Furthermore, with meticulous fine-tuning using instructional data, \dsins achieves better performance compared to the \textit{OpenAI GPT-3.5 Turbo} model in code-related tasks.
\end{itemize}

\section{Data Collection}
The training dataset of \dscoder is composed of 87\% source code, 10\% English code-related natural language corpus, and 3\% code-unrelated Chinese natural language corpus. The English corpus consists of materials from GitHub's Markdown and StackExchange\footnote{\url{https://stackexchange.com}}, which are used to enhance the model's understanding of code-related concepts and improve its ability to handle tasks like library usage and bug fixing. Meanwhile, the Chinese corpus consists of high-quality articles aimed at improving the model's proficiency in understanding the Chinese language.
In this section, we will provide an overview of how we construct the code training data. This process involves data crawling, rule-based filtering, dependency parsing, repository-level deduplication, and quality screening, as illustrated in Figure \ref{fig:data}. In the following, we will describe the data creation procedure step by step.

\begin{figure}[h]
    \centering
    \includegraphics[width=0.99\linewidth]{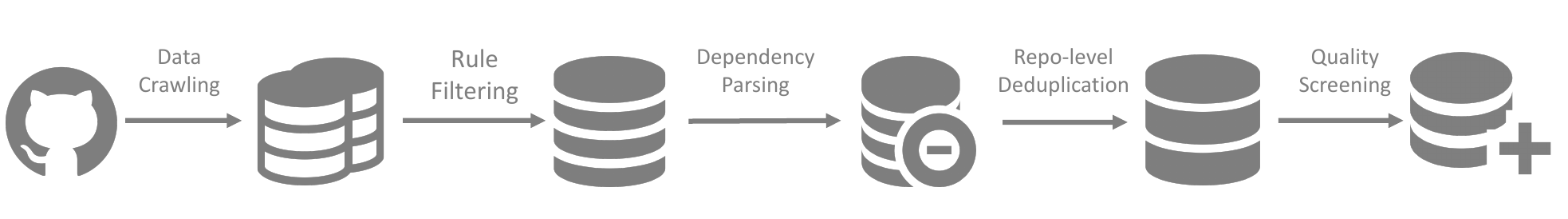}
    \caption{\centering The Procedure of Dataset Creation}
\label{fig:data}
\end{figure}

\subsection{GitHub Data Crawling and Filtering}
\label{sec:rule_filering}
We collect public repositories created before February 2023 on GitHub and retain only 87 programming languages, as listed in Table \ref{tab:programming_languages}. To reduce the amount of data to be processed, we apply filtering rules similar to those used in the StarCoder project ~\citep{li2023starcoder} to preliminarily filter out lower-quality code. By applying these filtering rules, we reduce the total amount of data to only 32.8\% of its original size. To make the paper self-contained, we briefly describe the filter rules used in the StarCoder Data project:

Firstly, we filter out files with an average line length exceeding 100 characters or a maximum line length surpassing 1000 characters. Additionally, we remove files with fewer than 25\% alphabetic characters.
Except for the XSLT programming language, we further filter out files where the string \texttt{"<?xml version="} appeared in the first 100 characters.
For HTML files, we consider the ratio of visible text to HTML code. We retain files where the visible text constitutes at least 20\% of the code and is no less than 100 characters.
For JSON and YAML files, which typically contain more data, we only keep files that have a character count ranging from 50 to 5000 characters. This effectively removes most data-heavy files.



\subsection{Dependency Parsing}
In previous works ~\citep{li2023starcoder,roziere2023code,nijkamp2022codegen,chen2021evaluating}, large language models for code are mainly pre-trained on file-level source code, which ignores the dependencies between different files in a project. However, in practical applications, such models struggle to effectively scale to handle entire project-level code scenarios.  Therefore, we will consider how to leverage the dependencies between files within the same repository in this step. Specifically, we first parse the dependencies between files and then arrange these files in an order that ensures the context each file relies on is placed before that file in the input sequence.
By aligning the files in accordance with their dependencies, our dataset more accurately represents real coding practices and structures. This enhanced alignment not only makes our dataset more relevant but also potentially increases the practicality and applicability of the model in handling project-level code scenarios. 
It's worth noting that we only consider the invocation relationships between files and use regular expressions to extract them, such as \textbf{"import"} in Python, \textbf{"using"} in C\#, and \textbf{"include"} in C.

\begin{algorithm}[!t]
\caption{Topological Sort for Dependency Analysis}
\label{alg:TopologicalSort}
\begin{algorithmic}[1]
\Procedure{TopologicalSort}{$files$}
    \State $graphs \gets \{\}$  \Comment{Initialize an empty adjacency list}
    \State $inDegree \gets \{\}$  \Comment{Initialize an empty dictionary for in-degrees}
    \For{\textbf{each} $file$ \textbf{in} $files$}
        \State $graphs[file] \gets []$ 
        \State $inDegree[file] \gets 0$ 
    \EndFor
    \State
    \For{\textbf{each} $fileA$ \textbf{in} $files$}
        \For{\textbf{each} $fileB$ \textbf{in} $files$}
            \If{\Call{HasDependency}{$fileA$, $fileB$}} \Comment{If fileA depends on fileB}
                \State $graphs[fileB].\text{append}(fileA)$  \Comment{Add edge from B to A}
                \State $inDegree[fileA] \gets inDegree[fileA] + 1$  \Comment{Increment in-degree of A}
            \EndIf
        \EndFor
    \EndFor
    \State
    \State $subgraphs \gets \text{getDisconnectedSubgraphs}(graphs)$ \Comment{Identify disconnected subgraphs}
    \State $allResults \gets []$
    \For{\textbf{each} $subgraph$ \textbf{in} $subgraphs$}
        \State $results \gets []$
        \While{$\text{length}(results) \neq \text{NumberOfNodes}(subgraph)$}
            \State $file \gets \text{argmin}(\{inDegree[file] \mid file \in subgraph \text{ and } file \notin results\})$
            \For{\textbf{each} $node$ \textbf{in} $graphs[file]$}
                \State $inDegree[node] \gets inDegree[node] - 1$
            \EndFor
            \State $results.\text{append}(file)$
        \EndWhile
        \State $allResults.\text{append}(results)$
    \EndFor
    \State
    \State \Return $allResults$
\EndProcedure
\end{algorithmic}
\end{algorithm}

The algorithm \ref{alg:TopologicalSort} describes a topological sort for dependency analysis on a list of files within the same project. Initially, it sets up two data structures: an empty adjacency list named \textbf{"graphs"} to represent dependencies between files and an empty dictionary called \textbf{"inDegree"} for storing the in-degrees of each file. The algorithm then iterates over each file pair to identify dependencies, updating \textbf{"graphs"} and \textbf{"inDegree"} accordingly. Next, it identifies any disconnected subgraphs within the overall dependency graph. For each subgraph, the algorithm employs a modified topological sort. Unlike the standard approach that selects nodes with zero in-degrees, this algorithm selects nodes with minimal in-degrees, which allows it to handle cycles within the graph. Selected nodes are added to a \textbf{"results"} list, and the in-degrees of their connected nodes are decreased. This process continues until a topologically sorted sequence is generated for each subgraph. The algorithm concludes by returning a list of these sorted sequences, and each sequence's files are concatenated to form a single training sample. To incorporate file path information, a comment indicating the file's path is added at the beginning of each file. This method ensures that the path information is preserved in the training data.

\subsection{Repo-Level Deduplication}
Recent studies have demonstrated the significant performance improvements that can be achieved by deduplicating training datasets for Large Language Models (LLMs).  \cite{lee2022deduplicating} have shown that language model training corpora often contain numerous near-duplicates, and the performance of LLMs can be enhanced by removing long repetitive substrings. \cite{kocetkov2022stack} have applied a near-deduplication method to training data, resulting in dramatic improvements, and they emphasize that near-deduplication is a crucial preprocessing step for achieving competitive performance on code benchmark tasks. In our dataset, we have also employed near-deduplication. However, there is a distinction in our approach compared to previous works. We perform deduplication at the repository level of code, rather than at the file level, as the latter approach may filter out certain files within a repository, potentially disrupting the structure of the repository. Specifically, we treat the concatenated code from the repository level as a single sample and apply the same near-deduplication algorithm to ensure the integrity of the repository structure.

\begin{table}[!bt]

\centering
\begin{small}

\resizebox{\linewidth}{!}{\begin{tabular}{l|c|c|c|l|c|c|c}
\toprule
Language & Size (GB) & Files (k) & Prop. (\%) & Language & Size (GB) & Files (k) & Prop. (\%) \\ 
\midrule
Ada & 0.91 & 126 & 0.11 & Literate Haskell & 0.16 & 20 & 0.02 \\
Agda & 0.26 & 59 & 0.03 & Lua & 0.82 & 138 & 0.10 \\
Alloy & 0.07 & 24 & 0.01 & Makefile & 0.92 & 460 & 0.12 \\
ANTLR & 0.19 & 38 & 0.02 & Maple & 0.03 & 6 & 0.00 \\
AppleScript & 0.03 & 17 & 0.00 & Mathematica & 0.82 & 10 & 0.10 \\
Assembly & 0.91 & 794 & 0.11 & MATLAB & 0.01 & 1 & 0.00 \\
Augeas & 0.00 & 1 & 0.00 & OCaml & 0.91 & 139 & 0.11 \\
AWK & 0.09 & 53 & 0.01 & Pascal & 0.79 & 470 & 0.10 \\
Batchfile & 0.92 & 859 & 0.12 & Perl & 0.81 & 148 & 0.10 \\
Bluespec & 0.10 & 15 & 0.01 & PHP & 58.92 & 40,627 & 7.38 \\
C & 28.64 & 27,111 & 3.59 & PowerShell & 0.91 & 236 & 0.11 \\
C\# & 58.56 & 53,739 & 7.34 & Prolog & 0.03 & 5 & 0.00 \\
Clojure & 0.90 & 295 & 0.11 & Protocol Buffer & 0.92 & 391 & 0.12 \\
CMake & 0.90 & 359 & 0.11 & Python & 120.68 & 75,188 & 15.12 \\
CoffeeScript & 0.92 & 361 & 0.12 & R & 0.92 & 158 & 0.11 \\
Common Lisp & 0.92 & 105 & 0.11 & Racket & 0.09 & 13 & 0.01 \\
C++ & 90.87 & 36,006 & 11.39 & RMarkdown & 6.83 & 1,606 & 0.86 \\
CSS & 5.63 & 11,638 & 0.71 & Ruby & 15.01 & 18,526 & 1.88 \\
CUDA & 0.91 & 115 & 0.11 & Rust & 0.61 & 692 & 0.08 \\
Dart & 0.89 & 264 & 0.11 & SAS & 0.92 & 70 & 0.11 \\
Dockerfile & 0.04 & 48 & 0.00 & Scala & 0.81 & 971 & 0.10 \\
Elixir & 0.91 & 549 & 0.11 & Scheme & 0.92 & 216 & 0.12 \\
Elm & 0.92 & 232 & 0.12 & Shell & 13.92 & 10,890 & 1.74 \\
Emacs Lisp & 0.91 & 148 & 0.11 & Smalltalk & 0.92 & 880 & 0.12 \\
Erlang & 0.92 & 145 & 0.12 & Solidity & 0.85 & 83 & 0.11 \\
F\# & 0.91 & 340 & 0.11 & Sparql & 0.10 & 88 & 0.01 \\
Fortran & 1.67 & 654 & 0.21 & SQL & 15.14 & 7,009 & 1.90 \\
GLSL & 0.92 & 296 & 0.11 & Stan & 0.20 & 41 & 0.03 \\
Go & 2.58 & 1,365 & 0.32 & Standard ML & 0.74 & 117 & 0.09 \\
Groovy & 0.89 & 340 & 0.11 & Stata & 0.91 & 122 & 0.11 \\
Haskell & 0.87 & 213 & 0.11 & SystemVerilog & 0.91 & 165 & 0.11 \\
HTML & 30.05 & 14,998 & 3.77 & TCL & 0.90 & 110 & 0.11 \\
Idris & 0.11 & 32 & 0.01 & Tcsh & 0.17 & 53 & 0.02 \\
Isabelle & 0.74 & 39 & 0.09 & Tex & 20.46 & 2,867 & 2.56 \\
Java & 148.66 & 134,367 & 18.63 & Thrift & 0.05 & 21 & 0.01 \\
Java Server Pages & 0.86 & 1072 & 0.11 & TypeScript & 60.62 & 62,432 & 7.60 \\
JavaScript & 53.84 & 71,895 & 6.75 & Verilog & 0.01 & 1 & 0.00 \\
JSON & 4.61 & 11956 & 0.58 & VHDL & 0.85 & 392 & 0.11 \\
Julia & 0.92 & 202 & 0.12 & Visual Basic & 0.75 & 73 & 0.09 \\
Jupyter Notebook & 14.38 & 2,555 & 1.80 & XSLT & 0.36 & 48 & 0.04 \\
Kotlin & 6.00 & 3,121 & 0.75 & Yacc & 0.72 & 67 & 0.09 \\
Lean & 0.52 & 68 & 0.07 & YAML & 0.74 & 890 & 0.09 \\
Literate Agda & 0.05 & 4 & 0.01 & Zig & 0.81 & 70 & 0.10 \\
Literate CoffeeScript & 0.01 & 3 & 0.00 & \bf{Total} & \bf{797.92} &\bf{603,173}& \bf{100.00}\\
\bottomrule
\end{tabular}}

\end{small}
\caption{\centering A summary of the cleaned training data for the selected programming languages.}
\label{tab:programming_languages}
\end{table}

\subsection{Quality Screening and Decontamination}
In addition to applying the filtering rules mentioned in Section \ref{sec:rule_filering}, we also employ a compiler and a quality model, combined with heuristic rules, to further filter out low-quality data. This includes code with syntax errors, poor readability, and low modularity. 
We provide the statistical summary of source code in Table \ref{tab:programming_languages}, which includes a total of 87 languages, detailing the disk size, number of files, and percentage for each language. The total data volume is 798 GB with 603 million files.
To ensure that our code training data is not contaminated by information from the test set, which may be present on GitHub, we've implemented an n-gram filtering process. This process involves the removal of any code segments that match specific criteria. Specifically, we filter out files containing docstrings, questions, and solutions from sources such as HumanEval ~\citep{chen2021evaluating}, MBPP ~\citep{austin2021program}, GSM8K \citep{gsm8k} and MATH \citep{hendrycks2021measuring}. For the filtering criteria, we apply the following rules: if a piece of code includes a 10-gram string identical to any in the test data, it is excluded from our training data. In cases where the test data comprises strings that are shorter than 10-grams but no less than 3-grams, we use an exact match approach for filtering.



\section{Training Policy}
\subsection{Training Strategy}
\subsubsection{Next Token Prediction} 
The first training objective for our model is known as \textit{next token prediction}. In this process, various files are concatenated to form a fixed-length entry. Then, these entries are used to train the model, enabling it to predict the subsequent token based on the provided context.

\subsubsection{Fill-in-the-Middle} 
The second training objective for our model is known as \textit{fill-in-the-middle.} In the code pre-training scenario, it is often necessary to generate corresponding inserted content based on the given context and subsequent text. Due to specific dependencies in a programming language, relying solely on next token prediction is insufficient to learn this fill-in-the-middle capability. Therefore, several approaches ~\citep{bavarian2022efficient,li2023starcoder} propose the pretraining method of Fill-in-the-Midlle (FIM). This approach involves randomly dividing the text into three parts, then shuffling the order of these parts and connecting them with special characters. This method aims to incorporate a fill-in-the-blank pretraining task during the training process. Within the FIM methodology, two distinct modes are employed: PSM (Prefix-Suffix-Middle) and SPM (Suffix-Prefix-Middle). In the PSM mode, the training corpus is organized in the sequence of $Prefix, Suffix, Middle$, aligning the text in a way that the middle segment is flanked by the prefix and suffix. Conversely, the SPM mode arranges the segments as $Suffix, Prefix, Middle$, presenting a different structural challenge. These modes are instrumental in enhancing the model's capability to handle various structural arrangements in code, providing a robust training framework for advanced code prediction tasks.
\begin{figure}[h]
    \centering
    \includegraphics[width=0.99\linewidth]{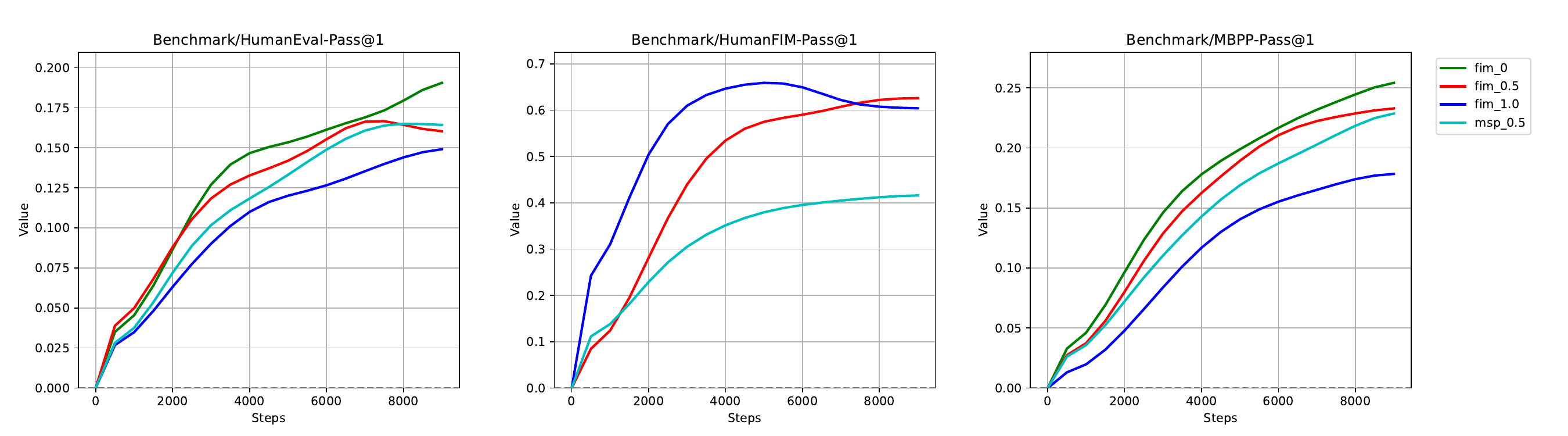}
    \caption{\centering The effectiveness of using FIM objective. }
\label{fig:fim}
\end{figure}

To determine the effectiveness of various hyperparameters within the FIM approach, we conducted a series of ablation experiments.

\noindent\textbf{Experiment Settings:} In this experiment, we employ \dsbase 1.3B as our model architecture. We focused on a Python subset from our training dataset to streamline the experimental process. Our primary objective was to assess the efficacy of the Fill-in-the-Middle (FIM) technique, utilizing the HumanEval-FIM benchmark \citep{fried2022incoder}. This benchmark specializes in a single-line FIM task for Python, in which one line of code from a HumanEval solution is randomly obscured, testing the model's proficiency in predicting the missing line. We hypothesize that the PSM mode may exhibit subtle differences compared to the traditional next-token prediction objective. This is primarily because PSM involves rearranging the order of the original text, potentially impacting the learning dynamics of the model. Therefore, we implement the PSM mode for FIM across four distinct configurations: 0\% FIM rate, 50\% FIM rate, 100\% FIM rate, and 50\% MSP rate. The Masked Span Prediction (MSP) strategy, initially introduced in T5 \citep{raffel2023exploring}, conceals multiple text spans and trains the model to reconstruct these segments. According to CodeGen2.5 \citep{nijkamp2023codegen2}, MSP may enhance FIM performance compared to PSM. Thus, we include this method in our comparative analysis.

\noindent\textbf{Results:} The outcomes of our experiment are illustrated in Figure \ref{fig:fim}. While the model demonstrates peak performance on the HumanEval-FIM with a 100\% FIM rate, this configuration also results in the weakest code completion capability. This indicates a trade-off between FIM and code completion abilities. Moreover, we observe that with a 50\% PSM rate, the model outperforms the MSP strategy. To achieve a balance between FIM efficiency and code completion proficiency, we ultimately choose the 50\% PSM rate as our preferred training policy.

In our implementation, we have introduced three sentinel tokens specifically for this task. For each code file, we initially divide its content into three segments, denoted as $f_{pre}$, $f_{middle}$, and $f_{suf}$. Using the PSM mode, we construct the training example as follows:
\begin{align}
\texttt{<｜fim\_start｜>}f_{pre}\texttt{<｜fim\_hole｜>}f_{suf}\texttt{<｜fim\_end｜>}f_{middle}\texttt{<|eos\_token|>} \nonumber
\end{align}
We implement the Fill-in-the-Middle (FIM) method at the document level before the packing process, as proposed in the original work by ~\cite{bavarian2022efficient}. This is done with an FIM rate of 0.5, following the PSM mode.

\subsection{Tokenizer}

For the tokenization process, we employ the HuggingFace Tokenizer library\footnote{\url{https://github.com/huggingface/tokenizers}} to train Byte Pair Encoding (BPE) tokenizers, as outlined in Sennrich et al. (2015) \citep{sennrich2015neural}, on a subset of our training corpus. Ultimately, we utilize a tokenizer configured with a vocabulary size of 32,000.
\subsection{Model Architecture}

We develop a range of models with varying parameters to cater to diverse applications, including models with 1.3B, 6.7B, and 33B parameters. These models are built upon the same framework as the DeepSeek Large Language Model (LLM) outlined by \cite{bi2024deepseek}. Each model is a decoder-only Transformer, incorporating Rotary Position Embedding (RoPE) as described by \cite{su2023roformer}. Notably, the DeepSeek 33B model integrates Grouped-Query-Attention (GQA) with a group size of 8, enhancing both training and inference efficiency. Additionally, we employ FlashAttention v2 \citep{dao2023flashattention2} to expedite the computation involved in the attention mechanism. The architectural details of our models are summarized in Table \ref{table:hparameter}. 

\subsection{Optimization}
Following DeepSeek LLM ~\citep{bi2024deepseek}, we use AdamW ~\citep{loshchilov2019decoupled} as the optimizer with $\beta_1$ and $\beta_2$ values of 0.9 and 0.95. We adapt batch sizes and learning rates by the scaling laws suggested in DeepSeek LLM. For the learning rate scheduling, we implement a three-stage policy, which includes 2000 warm-up steps, and set the final learning rate to 10\% of the initial rate. Notably, the learning rate at each stage is scaled down to $\sqrt{\frac{1}{10}}$ of the preceding stage's rate, following the guidelines established in DeepSeek LLM  \citep{bi2024deepseek}.
\begin{table}[h]
        \centering

        \begin{small}
	
		\centering

		\begin{tabular}{l|c|c|c}
			\toprule
			Hyperparameter&\dscoder 1.3B&\dscoder 6.7B&\dscoder 33B \\
               \midrule 
                Hidden Activation&SwiGLU&SwiGLU&SwiGLU \\
                Hidden size&2048&4096&7168 \\
                Intermediate size&5504&11008&19200 \\
                Hidden layers number&24&32&62\\
                Attention heads number&16&32&56\\
                Attention&Multi-head&Multi-head &Grouped-query (8) \\
                Batch Size &1024&2304&3840 \\
                Max Learning Rate&5.3e-4&4.2e-4& 3.5e-4\\
                \bottomrule 
		\end{tabular}
	\caption{\centering  Hyperparameters of \dscoder.}
        \label{table:hparameter}
        \end{small}
        
\end{table}
\subsection{Environments} Our experiments are conducted using the HAI-LLM \citep{haillm} framework, known for its efficiency and lightweight approach in training large language models. This framework incorporates a variety of parallelism strategies to optimize computational efficiency. These include tensor parallelism \citep{korthikanti2023reducing}, alongside ZeRO data parallelism \citep{rajbhandari2020zero} and PipeDream pipeline parallelism \citep{narayanan2019pipedream}. 
Our experiments utilize clusters outfitted with NVIDIA A100 and H800 GPUs. In the A100 cluster, each node is configured with 8 GPUs, interconnected in pairs using NVLink bridges. The H800 cluster is similarly arranged, with each node containing 8 GPUs. These GPUs are interconnected using a combination of NVLink and NVSwitch technologies, ensuring efficient data transfer within nodes. To facilitate seamless communication between nodes in both A100 and H800 clusters, we employ InfiniBand interconnects, known for their high throughput and low latency. This setup provides a robust and efficient infrastructure for our computational experiments.
\vspace{-0.1in}
\subsection{Long Context}

To enhance the capabilities of DeepSeek-Coder in handling extended contexts, particularly for scenarios like repository-level code processing, we have reconfigured the RoPE~\citep{su2023roformer} parameters to extend the default context window. Following previous practices~\citep{chen2023extending, superhot}, we employed a linear scaling strategy, increasing the scaling factor from $1$ to $4$ and altering the base frequency from $10000$ to $100000$. The model underwent an additional $1000$ steps of training, using a batch size of $512$ and a sequence length of $16$K. The learning rate was maintained as in the final pre-training phase. Theoretically, these modifications enable our model to process up to $64$K tokens in context. However, empirical observations suggest that the model delivers its most reliable outputs within a $16$K token range. Future research will continue to refine and evaluate the long-context adaptation methodology, aiming to further enhance DeepSeek-Coder's efficiency and user-friendliness in processing extended contexts.
\vspace{-0.1in}
\subsection{Instruction Tuning}
We develop \dsins by enhancing the \dsbase through instruction-based fine-tuning using high-quality data. This data comprises helpful and impartial human instructions, structured by the Alpaca Instruction format ~\citep{alpaca}. To demarcate each dialogue turn, we employed a unique delimiter token \texttt{<|EOT|>} to signify the conclusion of each segment. For training, we use a cosine schedule with 100 warm-up steps and an initial learning rate 1e-5. We also use a batch size of 4M tokens and 2B tokens in total. 

\begin{figure}[t]
    \centering
    \includegraphics[width=0.99\linewidth]{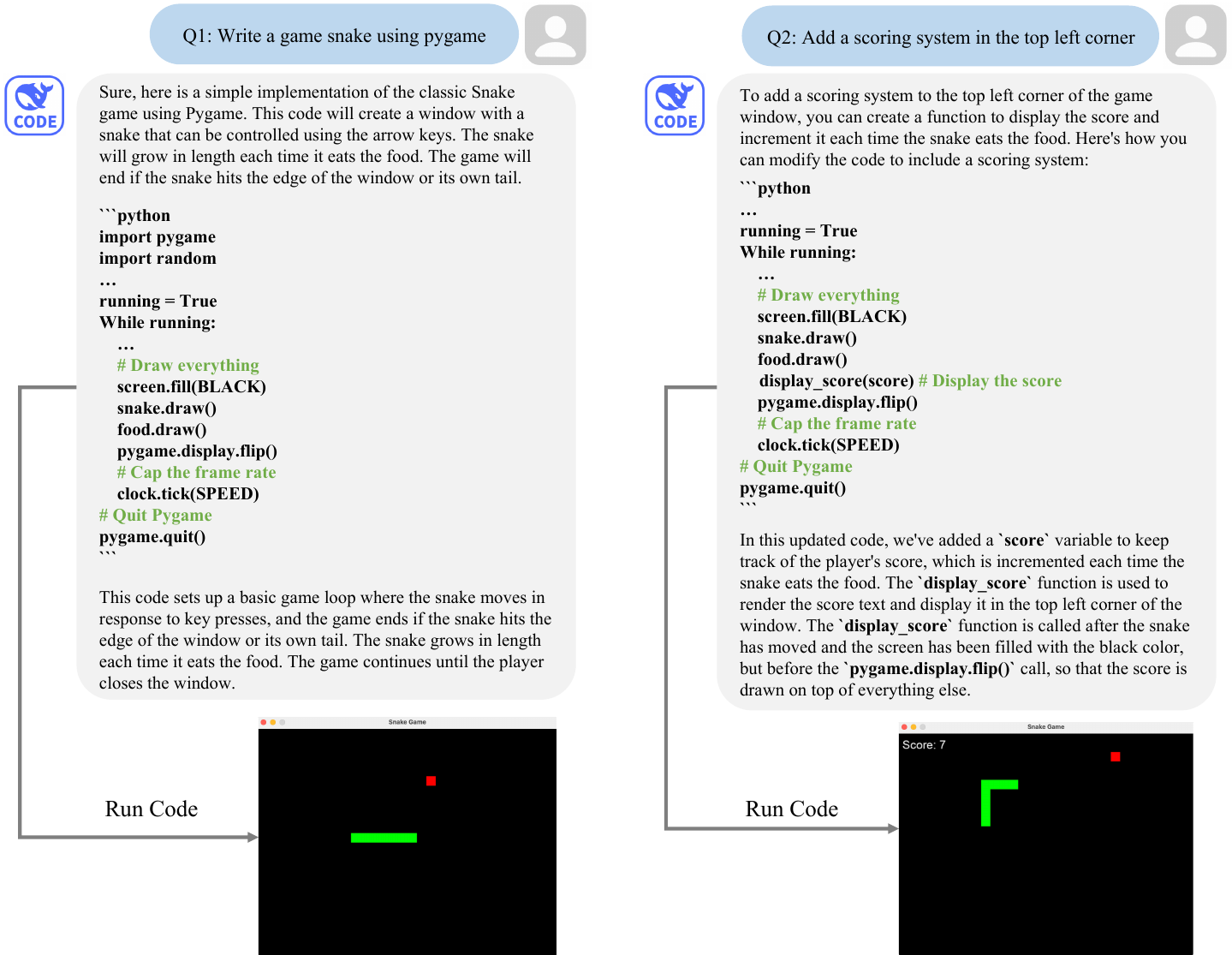}
    \caption{\centering An example of responses from \dsins 33B in a multi-turn setting.}
\label{fig:snake}
\end{figure}

An example of using \dsins 34B is depicted in Figure \ref{fig:snake}. This example is a multi-turn dialogue scenario for building a snake game. Initially, we ask the model to write a game snake using pygame. The model successfully creates a basic snake game that can run without bugs. To improve the game, we further request adding a scoring system in the top left corner. The model then introduces a \textbf{"score"} variable and a \textbf{"display\_score"} function, along with an explanation of how to integrate these features. This example illustrates \dsins's ability to provide complete solutions in multi-turn dialogue settings.
More cases can be found in the Appendix \ref{appendix_cases}.

\section{Experimental Results}
In this section, we evaluate \dscoder on four tasks, including code generation (\S \ref{sec:code_generation}),
FIM code completion (\S \ref{sec:code_understanding}), cross-file code completion (\S \ref{sec:code_completion}) and program-based math reasoning (\S \ref{sec:math_reasoning}). We compare \dscoder with the previous state-of-the-art large language models:
\begin{itemize}
    \item \textbf{CodeGeeX2} \citep{zheng2023codegeex} represents the second generation of the multilingual code generation model CodeGeeX. It is developed using the ChatGLM2 \citep{du2022glm} architecture and is enhanced with an extensive dataset of coding examples.
    \item \textbf{StarCoder} \citep{li2023starcoder} is a publicly accessible model with a substantial parameter count of 15 billion. It is specifically trained on a meticulously curated subset of the Stack dataset \citep{kocetkov2022stack}, covering 86 programming languages, ensuring its proficiency across a wide range of coding tasks.
    \item \textbf{CodeLlama} \citep{roziere2023code} encompasses a series of code-centric Large Language Models (LLMs) that are derivatives of LLaMA2 \citep{touvron2023llama}. Available in three sizes — 7B, 13B, and 34B — these models undergo continued training on a vast 500 billion token code corpus, building upon the foundational LLaMA2 architecture.
    \item \textbf{code-cushman-001}  \cite{chen2021evaluating} is a 12 billion parameter model developed by OpenAI and served as the initial model for Github Copilot.

    \item \textbf{GPT-3.5 and GPT-4} \citep{openai2023gpt4} are advanced generative AI models developed by OpenAI. While they are not explicitly trained for code generation, they also demonstrate notable performance in this domain. Their effectiveness in handling code generation tasks is largely attributed to their massive scale in terms of parameter count.
\end{itemize}

\subsection{Code Generation}
\label{sec:code_generation}

\paragraph{HumanEval and MBPP Benchmarks}
The HumanEval ~\citep{chen2021evaluating} and MBPP ~\citep{austin2021program} benchmarks are widely used for evaluating code LLMs. HumanEval consists of 164 hand-written Python problems that are validated using test cases to assess the code generated by a Code LLM in a zero-shot setting, while the MBPP benchmark includes 500 problems in a few-shot setting. To evaluate the model's multilingual capabilities, we expanded the Python problems of Humaneval Benchmark to seven additional commonly used programming languages, namely C++, Java, PHP, TypeScript (TS), C\#, Bash, and JavaScript (JS) \citep{cassano2023multipl}. For both benchmarks, We adopted a greedy search approach and re-implemented the baseline results using the same script and environment for fair comparison. 


\begin{table}[h]

        \begin{small}
	
		\centering
		\resizebox{\linewidth}{!}{\begin{tabular}{lc|c|c|c|c|c|c|c|c|c|c}
			\toprule
			Model&Size&Python&C++&Java&PHP&TS&C\#&Bash&JS&Avg &MBPP \\
               \midrule
               \multicolumn{12}{c}{Multilingual Base Models} \\
               \midrule
                code-cushman-001&12B&33.5\%&31.9\%&30.6\%&28.9\%&31.3\%&22.1\%&11.7\%&-&-&- \\
                CodeGeeX2&6B&36.0\%&29.2\%&25.9\%&23.6\%&20.8\%&29.7\%&6.3\%&24.8\%&24.5\%&36.2\%\\
                StarCoderBase&16B&31.7\%&31.1\%&28.5\%&25.4\%&34.0\%&34.8\%&8.9\%&29.8\%&28.0\%&42.8\% \\
                CodeLlama&7B&31.7\%&29.8\%&34.2\%&23.6\%&36.5\%&36.7\%&12.0\%&29.2\%&29.2\%&38.6\% \\
                CodeLlama&13B&36.0\%&37.9\%&38.0\%&34.2\%&45.2\%&43.0\%&16.5\%&32.3\%&35.4\%&48.4\% \\
                CodeLlama&34B&48.2\%&44.7\%&44.9\%&41.0\%&42.1\%&48.7\%&15.8\%&42.2\%&41.0\%&55.2\% \\
                \midrule
                \dsbase&1.3B&34.8\%&31.1\%&32.3\%&24.2\%&28.9\%&36.7\%&10.1\%&28.6\%&28.3\% &46.2\% \\
                \dsbase&6.7B&49.4\%&50.3\%&43.0\%&38.5\%&49.7\%&50.0\%&28.5\%&48.4\%&44.7\%&60.6\% \\
                \dsbase&33B&\bf{56.1\%}&\bf{58.4\%}&\bf{51.9\%}&\bf{44.1\%}&\bf{52.8\%}&\bf{51.3\%}&\bf{32.3\%}&\bf{55.3\%}&\bf{50.3\%}&\textbf{66.0\%} \\
                \midrule
               \multicolumn{12}{c}{Instruction-Tuned Models} \\
               \midrule
                GPT-3.5-Turbo&-&76.2\%&63.4\%&69.2\%&60.9\%&69.1\%&70.8\%&42.4\%&67.1\%&64.9\%&70.8\% \\
                GPT-4&-&\bf{84.1\%}&\bf{76.4\%}&\bf{81.6\%}&\bf{77.2\%}&\bf{77.4\%}&\bf{79.1\%}&\bf{58.2\%}&\bf{78.0\%}&\bf{76.5\%} &\bf{80.0\%}\\
                \midrule
                \dsins&1.3B&65.2\%&45.3\%&51.9\%&45.3\%&59.7\%&55.1\%&12.7\%&52.2\%&48.4\% &49.4\%\\
                \dsins&6.7B&78.6\%&63.4\%&68.4\%&68.9\%&67.2\%&72.8\%&36.7\%&72.7\%&66.1\%&65.4\% \\
                \dsins&33B&\bf{79.3\%}&\bf{68.9\%}&\bf{73.4\%}&\bf{72.7\%}&\bf{67.9\%}&\bf{74.1\%}&\bf{43.0\%}&\bf{73.9\%}& \bf{69.2\%}&\textbf{70.0\%} \\
		    \bottomrule
		\end{tabular}}
	\caption{\centering Performance of approaches on the Multilingual HumanEval and MBPP Benchmarks.}
        \label{table:result-humaneval}
        \end{small}
\end{table}

The results are presented in Table \ref{table:result-humaneval}. As we can see, \dsbase achieves state-of-the-art performance with an average accuracy of 50.3\% on HumanEval and 66.0\% on MBPP. In comparison to the similarly sized open-source model CodeLlama-Base 34B, our model has demonstrated a notable improvement of 9\% and 11\% in accuracy, respectively. It's worth noting that even our smaller model, \dsbase 6.7B, surpasses the performance of CodeLlama-Base 34B. After instruction fine-tuning, our model surpasses the closed-source GPT-3.5-Turbo model in HumanEval benchmark, significantly reducing the performance gap between OpenAI GPT-4 and open-source models.



\paragraph{DS-1000 Benchmark}
HumanEval and MBPP have a significant drawback in that they rely heavily on straightforward programming tasks that may not accurately represent the kind of code most programmers typically write. In contrast, the DS-1000 benchmark, as introduced in the work by  ~\cite{lai2023ds}, offers a comprehensive collection of 1,000 practical and realistic data science workflows across seven different libraries. This benchmark evaluates code generation by executing it against specific test cases. What sets DS-1000 apart is its categorization of problems based on the libraries involved, which encompass Matplotlib, NumPy, Pandas, SciPy, Scikit-Learn, PyTorch, and TensorFlow. The benchmark assesses the performance of base models in the code completion setting and we provide pass@1 results for each library, as well as overall score.

The results of DS-1000 benchmark are shown in Table \ref{table:result-ds-1000}. As can be seen from the table, the \dscoder model achieves relatively high accuracy in all libraries, demonstrating that our model is not only capable of generating good code but also of using libraries more accurately in real data science workflows.

\begin{table}[h]
        \label{}
        \begin{small}

		\centering
            \resizebox{\linewidth}{!}{
		\begin{tabular}{lc|c|c|c|c|c|c|c|c}
			\toprule
			Model&Size&Matplotlib&Numpy&Pandas&Pytorch&Scipy&Scikit-Learn&Tensorflow&Avg \\
               \midrule
                CodeGeeX2&6B& 38.7\%&26.8\%&14.4\%&11.8\%&19.8\%&27.0\%&17.8\%&22.9\% \\
                StarCoder-Base&16B&43.2\%&29.1\%&11.0\%&20.6\%&23.6\%&32.2\%&15.6\%&24.6\% \\
                CodeLlama-Base&7B&41.9\%&24.6\%&14.8\%&16.2\%&18.9\%&17.4\%&17.8\%&22.1\% \\
                CodeLlama-Base&13B&46.5\%&28.6\%&18.2\%&19.1\%&18.9\%&27.8\%&33.3\%&26.8\% \\
                CodeLlama-Base&34B&50.3\%&42.7\%&23.0\%&25.0\%&28.3\%&33.9\%&40.0\%&34.3\% \\
                \midrule
                \dsbase&1.3B&32.3\%&21.4\%&9.3\%&8.8\%&8.5\%&16.5\%&8.9\%&16.2\% \\
                \dsbase&6.7B&48.4\%&35.5\%&20.6\%&19.1\%&22.6\%&38.3\%&24.4\%&30.5\% \\
                \dsbase&33B&\bf{56.1\%}&\bf{49.6\%}&\bf{25.8\%}&\bf{36.8\%}&\bf{36.8\%}&\bf{40.0\%}&\bf{46.7\%}&\bf{40.2\%} \\
		    \bottomrule
		\end{tabular}}
	\caption{\centering  Performance of different approaches on the DS-1000-Tasks.}
 	\label{table:result-ds-1000}
        \end{small}
        \vspace{-0.1in}
\end{table}

\paragraph{LeetCode Contest Benchmark}
To further validate the model's capability in real-world programming problems, we construct the LeetCode Contest benchmark\footnote{We have published this benchmark in \url{https://github.com/deepseek-ai/DeepSeek-Coder/tree/main/Evaluation/LeetCode}.}. LeetCode\footnote{\url{https://leetcode.com/}} presents competition-level problems, offering significant challenges that test the model's problem understanding and code generation skills. We collected the latest problems from LeetCode Contests to prevent the appearance of both the problems or their solutions in our pre-training data. A total of 180 problems were collected from July 2023 to January 2024. For each problem, we collected 100 test cases to ensure the test coverage. We use the template "\{problem\_description\}\textbackslash nPlease complete the code below to solve the above problem:\textbackslash n\textasciigrave\textasciigrave\textasciigrave python\textbackslash n\{code\_template\}\textbackslash n\textasciigrave\textasciigrave\textasciigrave" to build the instruction prompt.

The evaluation results are shown in Table \ref{table:result-leetcode}.
In our evaluation, the \dscoder models demonstrate remarkable performance over current open-source coding models. Specifically, the \dsins 6.7B and 33B achieve Pass@1 scores of 19.4\% and 27.8\% respectively in this benchmark. This performance notably surpasses existing open-sourced models such as Code-Llama-33B. The \dsins 33B is the only open-sourced model that outperforms OpenAI's GPT-3.5-Turbo in this task. However, there remains a substantial performance gap when compared to the more advanced GPT-4-Turbo. 

\begin{table}[h]
        \centering

		\centering
            \begin{small}
		\begin{tabular}{lc|c|c|c|c}
			\toprule
			Model&Size&Easy (45)&Medium (91)&Hard (44)&Overall(180)\\
               \midrule 
                WizardCoder-V1.0 &15B&17.8\%&1.1\%&0.0\%&5.0\%\\
                CodeLlama-Instruct&34B&24.4\%&4.4\%&4.5\%&9.4\%\\
                Phind-CodeLlama-V2&34B&26.7\%&8.8\%&9.1\%&13.3\%\\
                \midrule
                GPT-3.5-Turbo & - &46.7\%& 15.4 \%&15.9\%& 23.3\%\\
                GPT-3.5-Turbo + CoT & - &42.2\%& 15.4\%&20.5\%& 23.3\%\\

                GPT-4-Turbo & - &\textbf{73.3\%}&31.9\%&25.0\%& 40.6\%\\
                GPT-4-Turbo + CoT & - &71.1\%&\textbf{35.2\%}&\textbf{25.0\%}& \textbf{41.8\%}\\
               \midrule
                \dsins       &1.3B& 22.2\%&1.1\%&4.5\%&7.2\%\\
                \dsins + CoT &1.3B& 22.2\%&  2.2\%& 2.3\%&  7.2\%\\
                \dsins       &6.7B& 44.4\%& 12.1\%& 9.1\%& 19.4\%\\
                \dsins + CoT&6.7B & 44.4\%& 17.6\%& 4.5\%& 21.1\%\\

                \dsins&33B&\textbf{57.8\%}&22.0\%&9.1\% &27.8\%\\
                \dsins + CoT&33B&53.3\%&\textbf{25.3\%}&\textbf{11.4\%} &\textbf{28.9\%}\\
                \bottomrule 
		\end{tabular}

	\caption{\centering  Performance of different models on the LeetCode Contest Benchmark.}
 \label{table:result-leetcode}
        \end{small}
\end{table}

Our analysis indicates that the implementation of Chain-of-Thought (CoT) prompting notably enhances the capabilities of \dsins models. This improvement becomes particularly evident in the more challenging subsets of tasks. By adding the directive, "You need first to write a step-by-step outline and then write the code." following the initial prompt, we have observed enhancements in performance. This observation leads us to believe that the process of first crafting detailed code descriptions assists the model in more effectively understanding and addressing the intricacies of logic and dependencies in coding tasks, particularly those of higher complexity. Therefore, we strongly recommend employing CoT prompting strategies when utilizing \dsins models for complex coding challenges. Such an approach promotes a more methodical and logical framework for problem-solving, potentially resulting in more precise and efficient outcomes in code generation tasks.

It is important to acknowledge that despite our diligent efforts to gather the most recent code questions for model testing, the possibility of data contamination cannot be entirely ruled out. We observed that the GPT-4-Turbo and \dscoder models achieved higher scores in the LeetCode Contest held in July and August. We encourage the research community to consider the potential issue of data contamination when evaluating models in future studies using our released LeetCode data.


\vspace{-0.1in}
\subsection{Fill-in-the-Middle Code Completion}
\label{sec:code_understanding}
\dscoder models are trained with a 0.5 FIM (Fill-In-the-Middle) rate during their pretraining phase. This specialized training strategy empowers the model to proficiently generate code by filling in blanks based on the surrounding context, both prefix and suffix, of the given code snippet. This capability is particularly advantageous in the realm of code completion tools. Several open-source models have emerged with similar capabilities. Notable among these are SantaCoder \citep{allal2023santacoder}, StarCoder \citep{li2023starcoder}, and CodeLlama \citep{roziere2023code}. These models have set a precedent in the field of code generation and completion. In evaluating the performance \dscoder models, we conducted a comparative analysis with the aforementioned models. The benchmark for this comparison was the Single-Line Infilling benchmarks, encompassing three different programming languages, as proposed by \citet{allal2023santacoder}. This benchmark uses the line exact match accuracy as the evaluation metric. 

\begin{table}[h]
		\centering
  \begin{small}
		\begin{tabular}{lc|c|c|c|c}
			\toprule
			Model&Size&python&java&javascript &Mean\\
               \midrule    
                SantaCoder&1.1B&44.0\%&62.0\%&74.0\%& 69.0\%\\
                StarCoder&16B&62.0\%&73.0\%&74.0\% &69.7\%\\
                CodeLlama-Base&7B&67.6\%&74.3\%&80.2\% &69.7\%\\
                CodeLlama-Base&13B&\textbf{68.3}\%&77.6\%&80.7\%&75.5\%\\
                \midrule
                DeepSeek-Coder-Base&1B&57.4\%&82.2\%&71.7\%& 70.4\%\\
                DeepSeek-Coder-Base&7B&66.6\%&\textbf{88.1}\%&79.7\%&80.7\%\\
                DeepSeek-Coder-Base&33B&65.4\%&86.6\%&\textbf{82.5}\% &\textbf{81.2}\%\\
                \bottomrule 
		\end{tabular}
	\caption{\centering Performance of different approaches on the FIM-Tasks.}        
	\label{table:compare-to-other-alg-fim}
        
            \end{small}
\end{table}

The evaluation results are shown in Table \ref{table:compare-to-other-alg-fim}. Despite being the smallest model with a capacity of 1.3 billion parameters, \dscoder outperforms its larger counterparts, StarCoder and CodeLlama, in these benchmarks. This superior performance can be attributed to the high quality of the pre-trained data utilized by DeepSeek-Coder. Furthermore, a notable trend observed is the correlation between the size of the model and its performance. As the model size increases, there is a corresponding and responsible enhancement in performance. This trend underscores the importance of model capacity in achieving higher accuracy in code completion tasks. Based on these findings, we recommend the deployment of the \dsbase 6.7B model in code completion tools. This recommendation is grounded in the model's demonstrated balance between efficiency and accuracy. The \dsbase 6.7B model, with its substantial parameter size, has proven to be highly effective in the context of code completion, making it an ideal choice for integrating advanced computational capabilities into coding environments.
\subsection{Cross-File Code Completion}
\label{sec:code_completion}
In this section, we will evaluate the performance of existing open-source models in cross-file code completion tasks. Unlike code generation discussed in the previous section, cross-file code completion requires the model to access and understand repositories that span multiple files with numerous cross-file dependencies.
We use CrossCodeEval~\citep{ding2023crosscodeeval} to evaluate the capabilities of currently available open-source code models of 7B scale in cross-file completion tasks. This dataset is constructed on a diverse set of real-world, open-sourced, permissively licensed repositories in four popular programming languages: Python, Java, TypeScript, and C\#. The dataset is specifically designed to strictly require cross-file context for accurate completion. Notably, this dataset was constructed from repositories created between March and June 2023, while our pre-training data only includes code created before February 2023, which ensures that this dataset was not present in our pre-training data, thus avoiding data leakage.

\begin{table}[h]
\centering
\begin{small}
\resizebox{\linewidth}{!}{\begin{tabular}{l l c c c c c c c c}
\toprule
\multirow{2}{*}{Model} &\multirow{2}{*}{Size} &\multicolumn{2}{c}{Python}&\multicolumn{2}{c}{Java}&\multicolumn{2}{c}{TypeScript}&\multicolumn{2}{c}{C\#} \\
\cmidrule(lr){3-4}\cmidrule(lr){5-6}\cmidrule(lr){7-8}\cmidrule(lr){9-10}&&EM&ES&EM&ES&EM&ES&EM&ES \\
\midrule
CodeGeex2 & 6B &  8.11\% & 59.55\% &7.34\%& 59.60\% & 6.14\% & 55.50\% & 1.70\%&51.66\%\\
\; + Retrieval &  & 10.73\% & 61.76\%& 10.10\% &59.56\% & 7.72\% & 55.17\% &4.64\% & 52.30\% \\
\midrule
StarCoder-Base & 7B & 6.68\% & 59.55\% & 8.65\% & 62.57\% & 5.01\% & 48.83\% & 4.75\% & 59.53\%\\
\; + Retrieval &  & 13.06\% & 64.24\%& 15.61\% &64.78\% & 7.54\% & 42.06\% &14.20\% & 65.03\% \\
\midrule
CodeLlama-Base & 7B & 7.32\% & 59.66\% & 9.68\% & 62.64\% & 8.19\% &58.50\% &4.07\%  & 59.19\% \\
\; + Retrieval &  & 13.02\% & 64.30\%  & 16.41\%  & \bf{64.64\%}  & 12.34\%  & 60.64\%  & 13.19\%  & 63.04\%  \\
\midrule
\dsbase & 6.7B & 9.53\% & 61.65\% & 10.80\% & 61.77\% & 9.59\% &60.17\% & 5.26\%  & 61.32\% \\
\; + Retrieval &  & \bf{16.14\%} & 66.51\%  & \bf{17.72\%}  & 63.18\%  & \bf{14.03\%}  & \bf{61.77\%}  & \bf{16.23\%}  & \bf{63.42\%}  \\
\; + Retrieval w/o Repo Pre-training &  & 16.02\%& \bf{66.65\%}  & 16.64\%  & 61.88\%  & 13.23\%  & 60.92\%  & 14.48\%  & 62.38\%  \\
\bottomrule

\end{tabular}}

\caption{\centering Performance of different models on cross-file code completion.}
\label{tab:code_completion_results}
\end{small}

\end{table}

In our evaluation of various models, we set the maximum sequence length to 2048 tokens, the maximum output length to 50 tokens, and a limit of 512 tokens for the cross-file context. For the cross-file context, we utilize the official BM25 search results provided by ~\cite{ding2023crosscodeeval}. Evaluation metrics include exact match and edit similarity. The results, presented in Table \ref{tab:code_completion_results}, demonstrate that \dscoder consistently outperforms other models in cross-file completion tasks across multiple languages, showcasing its superior practical application capabilities. When only utilizing file-level code corpus ({\bf{w/o Repo Pre-training}}) to pre-train \dscoder, we observe a decrease in performance in the Java, TypeScript, and C\# languages, indicating the effectiveness of the repository-level pre-training. 
\vspace{-0.1in}
\subsection{Program-based Math Reasoning}
\label{sec:math_reasoning}
Program-based math reasoning involves evaluating a model's ability to understand and solve mathematical problems through programming. This type of reasoning is critical in fields such as data analysis and scientific computing. To conduct this assessment, we utilize the Program-Aided Math Reasoning (PAL) method as outlined in \cite{gao2023pal}. This approach is applied across seven distinct benchmarks, each offering unique challenges and contexts. These benchmarks includes GSM8K~\citep{gsm8k}, MATH~\citep{hendrycks2021measuring}, GSM-Hard~\citep{gao2023pal}, SVAMP~\citep{patel2021nlp}, TabMWP~\citep{lu2022dynamic}, ASDiv~\citep{miao2020diverse} and MAWPS~\citep{gou2023tora}. In each of these benchmarks, the model is prompted to alternately describe a solution step in natural language and then execute that step with code. As seen in Table \ref{table:math-reasoning-results}, \dscoder models achieve a remarkable performance across all benchmarks, especially the 33B variant, which demonstrates the potential of using such models in applications that require complex mathematical computations and problem-solving abilities.

\begin{table}[h]

        \begin{small}

		\centering
            \resizebox{\linewidth}{!}{
		\begin{tabular}{lc|c|c|c|c|c|c|c|c}
            \toprule
            Model&Size&GSM8k&MATH&GSM-Hard&SVAMP&TabMWP&ASDiv&MAWPS&Avg \\
            \midrule
            \multicolumn{10}{c}{Multilingual Base Models} \\
            \midrule  
            CodeGeex-2&7B&22.2\%&9.7\%&23.6\%&39.0\%&44.6\%&48.5\%&66.0\%&36.2\% \\  
            StarCoder-Base&16B&23.4\%&10.3\%&23.0\%&42.4\%&45.0\%&54.9\%&81.1\%&40.0\% \\
            CodeLlama-Base&7B&31.2\%&12.1\%&30.2\%&54.2\%&52.9\%&59.6\%&82.6\%&46.1\% \\
            CodeLlama-Base&13B&43.1\%&14.4\%&40.2\%&59.2\%&60.3\%&63.6\%&85.3\%&52.3\% \\
            CodeLlama-Base&34B&58.2\%&21.2\%&51.8\%&70.3\%&69.8\%&70.7\%&91.8\%&62.0\% \\
            \midrule
            \dsbase&1.3B&14.6\%&16.8\%&14.5\%&36.7\%&30.0\%&48.2\%&62.3\%&31.9\% \\
            \dsbase&6.7B&43.2\%&19.2\%&40.3\%&58.4\%&67.9\%&67.2\%&87.0\%&54.7\% \\
            \dsbase&33B&\textbf{60.7\%}&\textbf{29.1\%}&\textbf{54.1\%}&\textbf{71.6\%}&\textbf{75.3\%}&\textbf{76.7\%}&\textbf{93.3\%}&\textbf{65.8\%} \\
            \bottomrule
		\end{tabular}}
	\caption{\centering  Performance of different approaches on the program-aid math reasoning tasks.}
 \label{table:math-reasoning-results}
        \end{small}
\end{table}

\section{Continue Pre-Training From General LLM}

To further enhance the natural language understanding and mathematical reasoning abilities of the \dscoder model, we perform additional pre-training from the general language model DeepSeek-LLM-7B Base \citep{bi2024deepseek} on 2 trillion tokens, resulting in  \dscoder-v1.5 7B. For this pre-training, we specifically use the data sources listed in Table \ref{tab:deepseek_coder_v1.5_source}. Unlike \dscoder, \dscoder-v1.5 employs solely a next token prediction objective with a 4K context length during its pre-training phase.

\begin{table}[h]
\centering
\begin{tabular}{l|c}
\toprule
Data Source & Percentage \\
\midrule
Source Code & 70\% \\
Markdown and StackExchange & 10\% \\
Natural language related to code & 7\% \\
Natural language related to math & 7\% \\
Bilingual (Chinese-English) natural language& 6\% \\
\bottomrule
\end{tabular}
\caption{\centering Data sources for \dscoder-v1.5 7B pre-training}
\label{tab:deepseek_coder_v1.5_source}
\end{table}

We conduct a comparison between \dscoder-v1.5 7B and \dscoder 6.7B, and re-run all benchmarks using our evaluation pipeline to ensure a fair comparison. We evaluate performance across a wide range of tasks, which can be categorized as follows:
\begin{itemize}
    \item \textbf{Programming}: This category includes evaluations in a multilingual setting using the HumanEval dataset by \cite{{chen2021evaluating}}, as well as evaluations in a Python setting using the MBPP dataset by \cite{austin2021program}
    \item \textbf{Math Reasoning}: We assess performance on math reasoning tasks using the GSM8K benchmark ~\citep{gsm8k} and the MATH ~\citep{hendrycks2021measuring} benchmark [4]. These tasks involve solving math problems by generating programs.
    \item \textbf{Natural Language} Our evaluation in natural language tasks includes  MMLU~\citep{hendrycks2020measuring}, BBH~\citep{suzgun2022challenging}, HellaSwag~\citep{zellers2019hellaswag}, Winogrande~\citep{sakaguchi2021winogrande}, and ARC-Challenge \citep{clark2018think} benchmarks.
\end{itemize}

The results for the Base and Instruct models are presented in Table \ref{tab:deepseek_coder_v1.5}. It is observed that the \dsbase-v1.5 model, despite a slight decrease in coding performance, shows marked improvements across most tasks when compared to the \dsbase model. In particular, in the Math Reasoning and Natural Language categories, \dsbase-v1.5 significantly outperforms its predecessor across all benchmarks, which also demonstrates significant improvements in its mathematical reasoning and natural language processing capabilities.

\begingroup
\setlength{\tabcolsep}{3pt} 
\renewcommand{\arraystretch}{1} 
\begin{table*}[htb]
    \centering
    \resizebox{\linewidth}{!}{
\begin{tabular}{llccccccccc} 
\toprule
&& \multicolumn{2}{c}{Programming} & \multicolumn{2}{c}{Math Reasoning} & \multicolumn{5}{c}{Natural Language}   \\            
\cmidrule(lr){3-4} \cmidrule(lr){5-6} \cmidrule(lr){7-11}
Models &Size &HumanEval & MBPP  & GSM8K  & MATH &MMLU       & BBH         & HellaSwag & WinoG & ARC-C   \\ 
\midrule
\dsbase & 6.7B  & \bf{44.7\%} & \bf{60.6\%} &43.2\%&19.2\% &36.6\% &44.3\% & 53.8\% & 57.1\% & 32.5\% \\
\dsbase-v1.5 & 6.9B  & 43.2\% & 60.4\% &\bf{62.4\%} &\bf{24.7\%} & \bf{49.1\%} & \bf{55.2\%} &\bf{69.9\%} & \bf{63.8\%} & \bf{47.2\%} \\
\midrule
\dsins & 6.7B  & \bf{66.1\%} & \bf{65.4\%}&62.8\%&28.6\% & 37.2\% &46.9\%& 55.0\% & 57.6\% & 37.4\% \\
\dsins-v1.5 & 6.9B &64.1\%
 &64.6\%& \bf{72.6\%} &\bf{34.1\%} &\bf{49.5\%}&\bf{53.3\%} & \bf{72.2\%} & \bf{63.4\%}& \bf{48.1\%} \\

\bottomrule
\end{tabular}
    }
    \caption{
    \centering Comparative analysis of performance between \dsbase and \dsbase-v1.5. Math tasks are solved through programming.
    }
    \label{tab:deepseek_coder_v1.5}
\end{table*}
\endgroup
\section{Conclusion}

In this technical report, we introduce a series of specialized Large Language Models (LLMs) for coding, named \dscoder, available in three distinct scales: 1.3B, 6.7B, and 33B parameters. These models are uniquely trained on a meticulously curated project-level code corpus, utilizing a "fill-in-the-blank" pre-training objective to enhance code infilling capabilities. A significant advancement is the extension of the models' context window to 16,384 tokens, thereby greatly improving their effectiveness in handling extensive code generation tasks. Our evaluations reveal that the most advanced model in our series, \dsbase 33B surpasses existing open-source code models across a variety of standard tests. Impressively, the \dsbase 6.7B model, despite its smaller scale, delivers performance on par with the 34B parameter CodeLlama, a testament to the high quality of our pretraining corpus.

To augment the zero-shot instruction capabilities of the \dsbase models, we have fine-tuned them with high-quality instructional data. This has led to the \dsins 33B model outperforming OpenAI's GPT-3.5 Turbo in a range of coding-related tasks, showcasing its exceptional proficiency in code generation and understanding.

To further improve the natural language understanding capabilities of the \dsbase models, we have conducted additional pretraining based on the DeepSeek-LLM 7B checkpoint. This additional training involved processing a diverse dataset comprising 2 billion tokens, including natural language, code, and mathematical data. The result is the creation of a new and improved code model, \dscoder-v1.5. Our observations indicate that \dscoder-v1.5 not only maintains its predecessor's high-level coding performance but also exhibits enhanced natural language comprehension. This advancement underscores our belief that the most effective code-focused Large Language Models (LLMs) are those built upon robust general LLMs. The reason is evident: to effectively interpret and execute coding tasks, these models must also possess a deep understanding of human instructions, which often come in various forms of natural language. Looking ahead, our commitment is to develop and openly share even more powerful code-focused LLMs based on larger-scale general LLMs.

\section*{Acknowledgements}
We would like to express our gratitude to  Bo Liu, Chengqi Deng, Chong Ruan, Damai Dai, Jiashi Li, Kang Guan, Mingchuan Zhang, Panpan Huang, Shuiping Yu, Shirong Ma, Yaofeng Sun, Yishi Piao, Zhihong Shao, and Zhewen Hao for their invaluable discussions and assistance during training \dscoder models.

\bibliography{main}

\newpage
\appendix


\section{Cases of Chatting with \dsins}
\label{appendix_cases}
We will present two cases of interactions with \dsins, with one involving a multi-turn conversation about creating a database and performing data analysis, and the other centered around using a model to solve a sample problem from LeetCode.

\begin{figure}[h]
    \centering
    \includegraphics[width=0.99\linewidth]{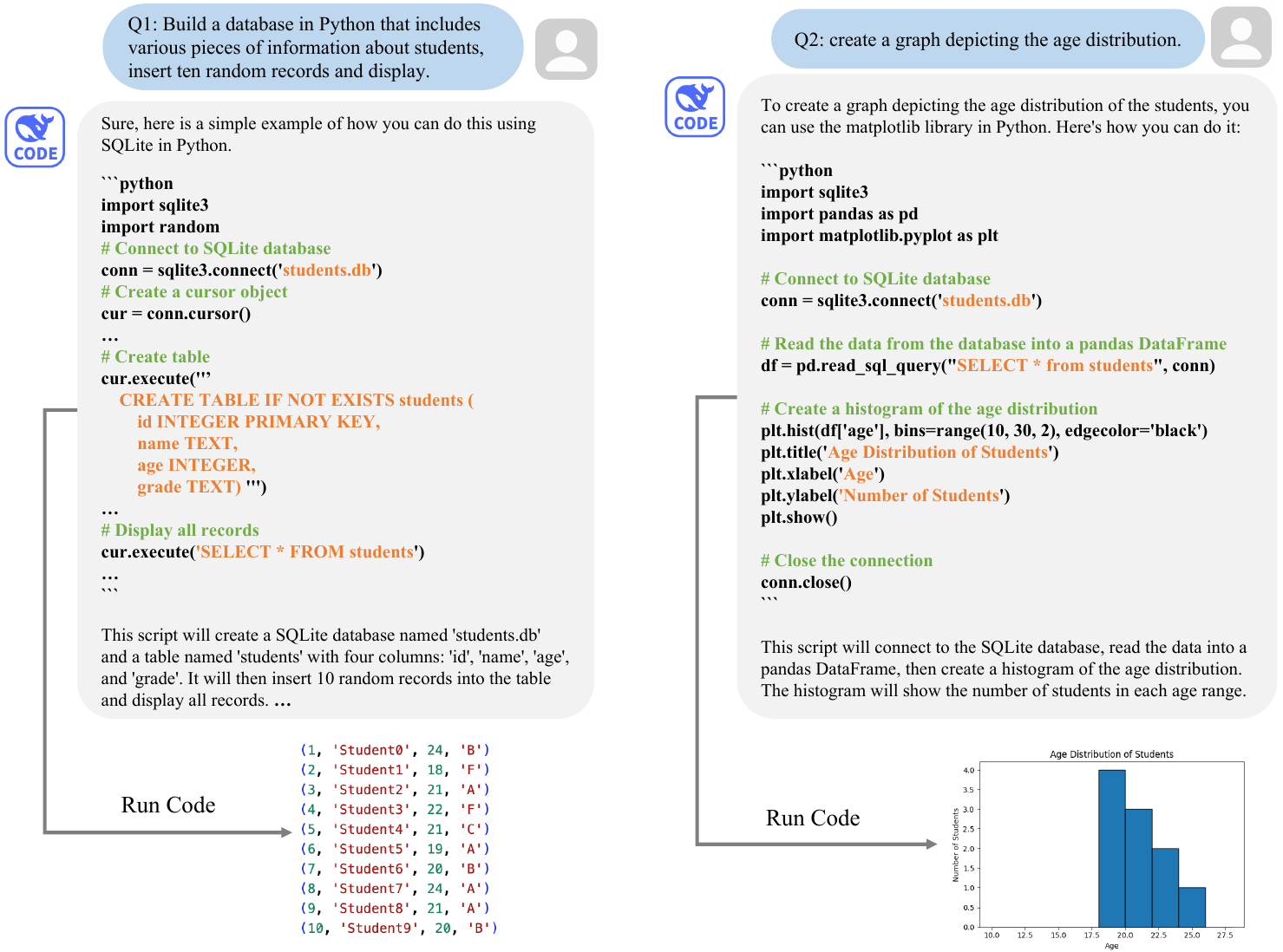}
    \caption{\centering An example of building database and data analysis.}
\label{fig:sql}
\end{figure}

\begin{figure}[t]
    \centering
    \includegraphics[width=0.99\linewidth]{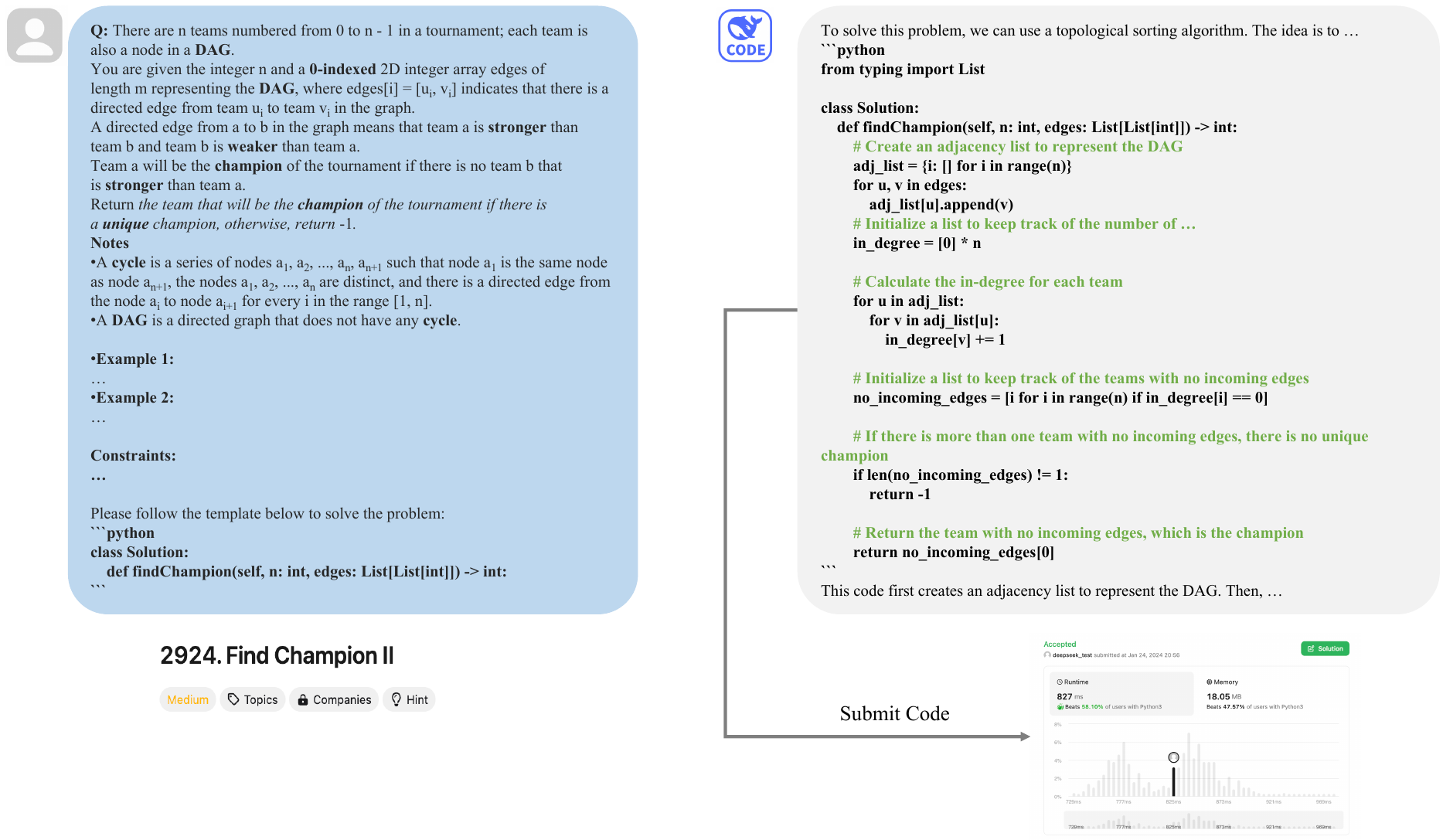}
    \caption{\centering An example of solving LeetCode Problem.}
\label{fig:leetcode}
\end{figure}

In the first scenario, depicted in Figure \ref{fig:sql}, we instruct the model to build a student database using Python and randomly insert 10 pieces of information. Subsequently, in the second round of the conversation, we continue to ask the model by analyzing the age distribution of the students. From Figure \ref{fig:sql}, it's evident that the model can generate bug-free and comprehensive code, accompanied by explanatory details. In the second scenario, as illustrated in Figure \ref{fig:leetcode}, we further assess the model's capabilities by testing it on an out-of-domain LeetCode contest problem. This particular problem was released in November 2023, after our data collection, and thus, isn't part of our model's training data. The results show that our model excels at solving problems that extend beyond its training distribution.

\newpage
\section{Benchmark curves during training of \dsbase}

\begin{figure}[h]
    \centering
    \includegraphics[width=0.95\linewidth]{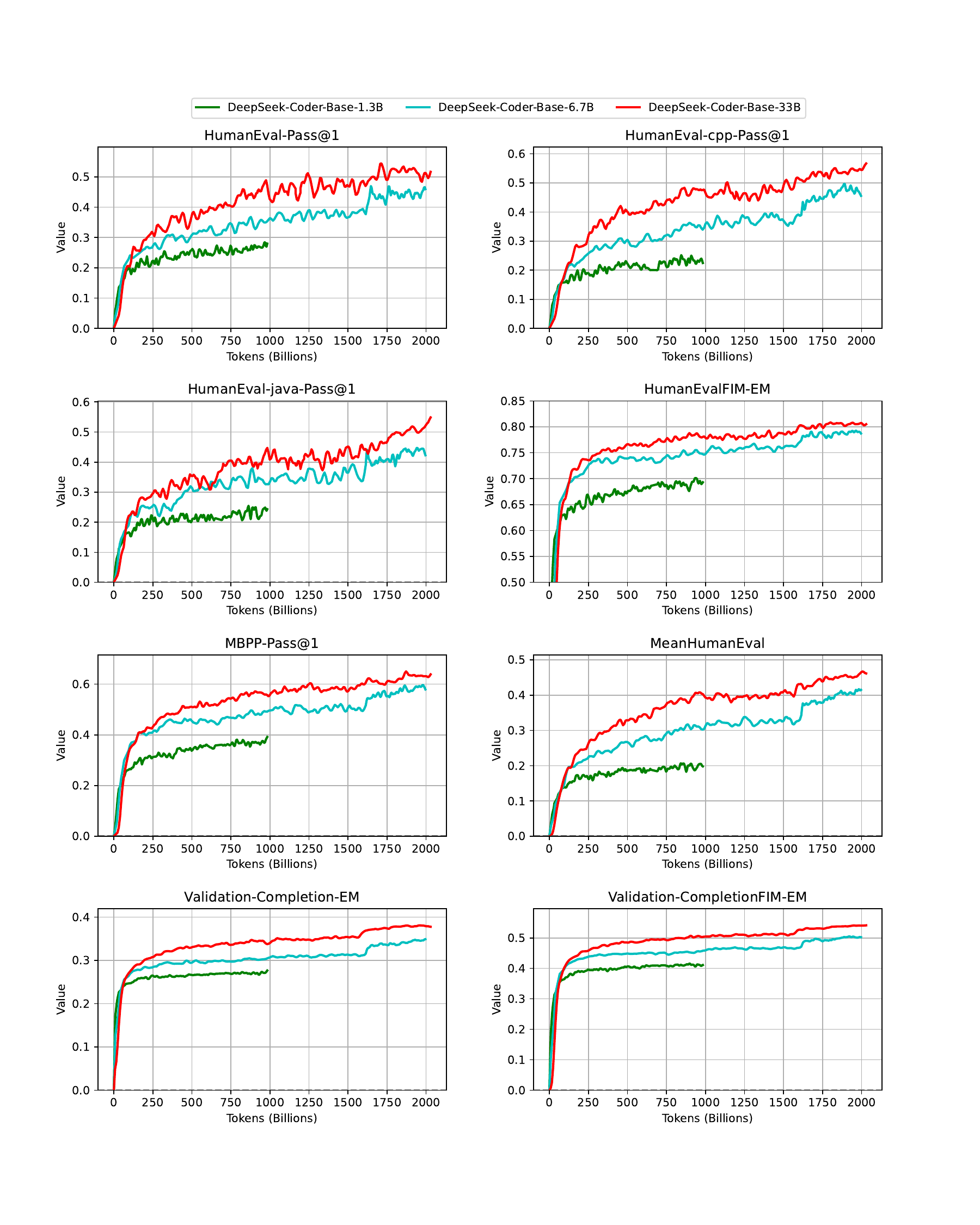}
    \caption{\centering Benchmark curves during training of \dsbase.}
\label{fig:curve}
\end{figure}

In Figure \ref{fig:curve}, we present the benchmark curves illustrating the performance of \dsbase models during their training phase. For validation, a carefully curated subset of the training corpus was employed, consisting of 8,000 code files. This subset was deliberately chosen to ensure a diverse and representative sample, critical for an accurate assessment of the models' capabilities. The performance metrics of these models are specifically detailed in the final two sub-figures of Figure \ref{fig:curve}, offering a clear visual representation of their efficacy throughout the training process.


\setcounter{figure}{0}
\makeatletter 
\renewcommand{\thefigure}{A\@arabic\c@figure}
\makeatother

\setcounter{table}{0}
\makeatletter 
\renewcommand{\thetable}{A\@arabic\c@table}
\makeatother

\end{CJK*}
\end{document}

%% file: commands.tex
\renewcommand{\phi}{\varphi}

\renewcommand{\epsilon}{\varepsilon}
\renewcommand{\imath}{\mathrm{i}}

\newlength{\restsubwidth}
\newlength{\restsubheight}
\newlength{\restsubmoreheight}
\newcommand{\rest}[2]{%
        \settowidth{\restsubwidth}{\ensuremath{#2}}
        \settoheight{\restsubheight}{\ensuremath{{}_{#2}}}
        \ensuremath{{#1\hskip 0.5pt}_{\vrule\kern2pt\parbox[b][%
        4pt][b]{\the\restsubwidth}{%
                        \ensuremath{{}_{#2}}}}}
        }

%% file: main.bbl
\begin{thebibliography}{45}
\providecommand{\natexlab}[1]{#1}
\providecommand{\url}[1]{\texttt{#1}}
\expandafter\ifx\csname urlstyle\endcsname\relax
  \providecommand{\doi}[1]{doi: #1}\else
  \providecommand{\doi}{doi: \begingroup \urlstyle{rm}\Url}\fi

\bibitem[Allal et~al.(2023)Allal, Li, Kocetkov, Mou, Akiki, Ferrandis, Muennighoff, Mishra, Gu, Dey, et~al.]{allal2023santacoder}
L.~B. Allal, R.~Li, D.~Kocetkov, C.~Mou, C.~Akiki, C.~M. Ferrandis, N.~Muennighoff, M.~Mishra, A.~Gu, M.~Dey, et~al.
\newblock Santacoder: don't reach for the stars!
\newblock \emph{arXiv preprint arXiv:2301.03988}, 2023.

\bibitem[Austin et~al.(2021)Austin, Odena, Nye, Bosma, Michalewski, Dohan, Jiang, Cai, Terry, Le, and Sutton]{austin2021program}
J.~Austin, A.~Odena, M.~Nye, M.~Bosma, H.~Michalewski, D.~Dohan, E.~Jiang, C.~Cai, M.~Terry, Q.~Le, and C.~Sutton.
\newblock Program synthesis with large language models, 2021.

\bibitem[Bavarian et~al.(2022)Bavarian, Jun, Tezak, Schulman, McLeavey, Tworek, and Chen]{bavarian2022efficient}
M.~Bavarian, H.~Jun, N.~Tezak, J.~Schulman, C.~McLeavey, J.~Tworek, and M.~Chen.
\newblock Efficient training of language models to fill in the middle.
\newblock \emph{arXiv preprint arXiv:2207.14255}, 2022.

\bibitem[Cassano et~al.(2023)Cassano, Gouwar, Nguyen, Nguyen, Phipps-Costin, Pinckney, Yee, Zi, Anderson, Feldman, et~al.]{cassano2023multipl}
F.~Cassano, J.~Gouwar, D.~Nguyen, S.~Nguyen, L.~Phipps-Costin, D.~Pinckney, M.-H. Yee, Y.~Zi, C.~J. Anderson, M.~Q. Feldman, et~al.
\newblock Multipl-e: a scalable and polyglot approach to benchmarking neural code generation.
\newblock \emph{IEEE Transactions on Software Engineering}, 2023.

\bibitem[Chen et~al.(2021)Chen, Tworek, Jun, Yuan, Pinto, Kaplan, Edwards, Burda, Joseph, Brockman, et~al.]{chen2021evaluating}
M.~Chen, J.~Tworek, H.~Jun, Q.~Yuan, H.~P. d.~O. Pinto, J.~Kaplan, H.~Edwards, Y.~Burda, N.~Joseph, G.~Brockman, et~al.
\newblock Evaluating large language models trained on code.
\newblock \emph{arXiv preprint arXiv:2107.03374}, 2021.

\bibitem[Chen et~al.(2023)Chen, Wong, Chen, and Tian]{chen2023extending}
S.~Chen, S.~Wong, L.~Chen, and Y.~Tian.
\newblock Extending context window of large language models via positional interpolation.
\newblock \emph{arXiv preprint arXiv:2306.15595}, 2023.

\bibitem[Clark et~al.(2018)Clark, Cowhey, Etzioni, Khot, Sabharwal, Schoenick, and Tafjord]{clark2018think}
P.~Clark, I.~Cowhey, O.~Etzioni, T.~Khot, A.~Sabharwal, C.~Schoenick, and O.~Tafjord.
\newblock Think you have solved question answering? try arc, the ai2 reasoning challenge.
\newblock \emph{arXiv preprint arXiv:1803.05457}, 2018.

\bibitem[Cobbe et~al.(2021)Cobbe, Kosaraju, Bavarian, Chen, Jun, Kaiser, Plappert, Tworek, Hilton, Nakano, et~al.]{gsm8k}
K.~Cobbe, V.~Kosaraju, M.~Bavarian, M.~Chen, H.~Jun, L.~Kaiser, M.~Plappert, J.~Tworek, J.~Hilton, R.~Nakano, et~al.
\newblock Training verifiers to solve math word problems.
\newblock \emph{arXiv preprint arXiv:2110.14168}, 2021.

\bibitem[Dao(2023)]{dao2023flashattention2}
T.~Dao.
\newblock Flashattention-2: Faster attention with better parallelism and work partitioning, 2023.

\bibitem[DeepSeek-AI(2024)]{bi2024deepseek}
DeepSeek-AI.
\newblock Deepseek llm: Scaling open-source language models with longtermism.
\newblock \emph{arXiv preprint arXiv:2401.02954}, 2024.

\bibitem[Ding et~al.(2023)Ding, Wang, Ahmad, Ding, Tan, Jain, Ramanathan, Nallapati, Bhatia, Roth, et~al.]{ding2023crosscodeeval}
Y.~Ding, Z.~Wang, W.~U. Ahmad, H.~Ding, M.~Tan, N.~Jain, M.~K. Ramanathan, R.~Nallapati, P.~Bhatia, D.~Roth, et~al.
\newblock Crosscodeeval: A diverse and multilingual benchmark for cross-file code completion.
\newblock In \emph{Thirty-seventh Conference on Neural Information Processing Systems Datasets and Benchmarks Track}, 2023.

\bibitem[Du et~al.(2022)Du, Qian, Liu, Ding, Qiu, Yang, and Tang]{du2022glm}
Z.~Du, Y.~Qian, X.~Liu, M.~Ding, J.~Qiu, Z.~Yang, and J.~Tang.
\newblock Glm: General language model pretraining with autoregressive blank infilling.
\newblock In \emph{Proceedings of the 60th Annual Meeting of the Association for Computational Linguistics (Volume 1: Long Papers)}, pages 320--335, 2022.

\bibitem[Fried et~al.(2022)Fried, Aghajanyan, Lin, Wang, Wallace, Shi, Zhong, Yih, Zettlemoyer, and Lewis]{fried2022incoder}
D.~Fried, A.~Aghajanyan, J.~Lin, S.~Wang, E.~Wallace, F.~Shi, R.~Zhong, W.-t. Yih, L.~Zettlemoyer, and M.~Lewis.
\newblock Incoder: A generative model for code infilling and synthesis.
\newblock \emph{arXiv preprint arXiv:2204.05999}, 2022.

\bibitem[Gao et~al.(2023)Gao, Madaan, Zhou, Alon, Liu, Yang, Callan, and Neubig]{gao2023pal}
L.~Gao, A.~Madaan, S.~Zhou, U.~Alon, P.~Liu, Y.~Yang, J.~Callan, and G.~Neubig.
\newblock Pal: Program-aided language models.
\newblock In \emph{International Conference on Machine Learning}, pages 10764--10799. PMLR, 2023.

\bibitem[Gemini~Team(2023)]{gemini}
G.~Gemini~Team.
\newblock Gemini: A family of highly capable multimodal models, 2023.
\newblock URL \url{https://goo.gle/GeminiPaper}.

\bibitem[Gou et~al.(2023)Gou, Shao, Gong, Yang, Huang, Duan, Chen, et~al.]{gou2023tora}
Z.~Gou, Z.~Shao, Y.~Gong, Y.~Yang, M.~Huang, N.~Duan, W.~Chen, et~al.
\newblock Tora: A tool-integrated reasoning agent for mathematical problem solving.
\newblock \emph{arXiv preprint arXiv:2309.17452}, 2023.

\bibitem[Hendrycks et~al.(2020)Hendrycks, Burns, Basart, Zou, Mazeika, Song, and Steinhardt]{hendrycks2020measuring}
D.~Hendrycks, C.~Burns, S.~Basart, A.~Zou, M.~Mazeika, D.~Song, and J.~Steinhardt.
\newblock Measuring massive multitask language understanding.
\newblock In \emph{International Conference on Learning Representations}, 2020.

\bibitem[Hendrycks et~al.(2021)Hendrycks, Burns, Kadavath, Arora, Basart, Tang, Song, and Steinhardt]{hendrycks2021measuring}
D.~Hendrycks, C.~Burns, S.~Kadavath, A.~Arora, S.~Basart, E.~Tang, D.~Song, and J.~Steinhardt.
\newblock Measuring mathematical problem solving with the math dataset.
\newblock \emph{arXiv preprint arXiv:2103.03874}, 2021.

\bibitem[High-Flyer(2023)]{haillm}
High-Flyer.
\newblock Hai-llm: An efficient and lightweight tool for training large models.
\newblock 2023.
\newblock URL \url{https://www.high-flyer.cn/en/blog/hai-llm}.

\bibitem[kaiokendev(2023)]{superhot}
kaiokendev.
\newblock Things i'm learning while training superhot.
\newblock \url{https://kaiokendev.github.io/til#extending-context-to-8k}, 2023.

\bibitem[Kocetkov et~al.(2022)Kocetkov, Li, Jia, Mou, Jernite, Mitchell, Ferrandis, Hughes, Wolf, Bahdanau, et~al.]{kocetkov2022stack}
D.~Kocetkov, R.~Li, L.~Jia, C.~Mou, Y.~Jernite, M.~Mitchell, C.~M. Ferrandis, S.~Hughes, T.~Wolf, D.~Bahdanau, et~al.
\newblock The stack: 3 tb of permissively licensed source code.
\newblock \emph{Transactions on Machine Learning Research}, 2022.

\bibitem[Korthikanti et~al.(2023)Korthikanti, Casper, Lym, McAfee, Andersch, Shoeybi, and Catanzaro]{korthikanti2023reducing}
V.~A. Korthikanti, J.~Casper, S.~Lym, L.~McAfee, M.~Andersch, M.~Shoeybi, and B.~Catanzaro.
\newblock Reducing activation recomputation in large transformer models.
\newblock \emph{Proceedings of Machine Learning and Systems}, 5, 2023.

\bibitem[Lai et~al.(2023)Lai, Li, Wang, Zhang, Zhong, Zettlemoyer, Yih, Fried, Wang, and Yu]{lai2023ds}
Y.~Lai, C.~Li, Y.~Wang, T.~Zhang, R.~Zhong, L.~Zettlemoyer, W.-t. Yih, D.~Fried, S.~Wang, and T.~Yu.
\newblock Ds-1000: A natural and reliable benchmark for data science code generation.
\newblock In \emph{International Conference on Machine Learning}, pages 18319--18345. PMLR, 2023.

\bibitem[Lee et~al.(2022)Lee, Ippolito, Nystrom, Zhang, Eck, Callison-Burch, and Carlini]{lee2022deduplicating}
K.~Lee, D.~Ippolito, A.~Nystrom, C.~Zhang, D.~Eck, C.~Callison-Burch, and N.~Carlini.
\newblock Deduplicating training data makes language models better.
\newblock In \emph{Proceedings of the 60th Annual Meeting of the Association for Computational Linguistics (Volume 1: Long Papers)}, pages 8424--8445, 2022.

\bibitem[Li et~al.(2023)Li, Allal, Zi, Muennighoff, Kocetkov, Mou, Marone, Akiki, Li, Chim, et~al.]{li2023starcoder}
R.~Li, L.~B. Allal, Y.~Zi, N.~Muennighoff, D.~Kocetkov, C.~Mou, M.~Marone, C.~Akiki, J.~Li, J.~Chim, et~al.
\newblock Starcoder: may the source be with you!
\newblock \emph{arXiv preprint arXiv:2305.06161}, 2023.

\bibitem[Loshchilov and Hutter(2019)]{loshchilov2019decoupled}
I.~Loshchilov and F.~Hutter.
\newblock Decoupled weight decay regularization, 2019.

\bibitem[Lu et~al.(2022)Lu, Qiu, Chang, Wu, Zhu, Rajpurohit, Clark, and Kalyan]{lu2022dynamic}
P.~Lu, L.~Qiu, K.-W. Chang, Y.~N. Wu, S.-C. Zhu, T.~Rajpurohit, P.~Clark, and A.~Kalyan.
\newblock Dynamic prompt learning via policy gradient for semi-structured mathematical reasoning.
\newblock In \emph{The Eleventh International Conference on Learning Representations}, 2022.

\bibitem[Miao et~al.(2020)Miao, Liang, and Su]{miao2020diverse}
S.-Y. Miao, C.-C. Liang, and K.-Y. Su.
\newblock A diverse corpus for evaluating and developing english math word problem solvers.
\newblock In \emph{Proceedings of the 58th Annual Meeting of the Association for Computational Linguistics}, pages 975--984, 2020.

\bibitem[Narayanan et~al.(2019)Narayanan, Harlap, Phanishayee, Seshadri, Devanur, Ganger, Gibbons, and Zaharia]{narayanan2019pipedream}
D.~Narayanan, A.~Harlap, A.~Phanishayee, V.~Seshadri, N.~R. Devanur, G.~R. Ganger, P.~B. Gibbons, and M.~Zaharia.
\newblock Pipedream: Generalized pipeline parallelism for dnn training.
\newblock In \emph{Proceedings of the 27th ACM Symposium on Operating Systems Principles}, pages 1--15, 2019.

\bibitem[Nijkamp et~al.(2022)Nijkamp, Pang, Hayashi, Tu, Wang, Zhou, Savarese, and Xiong]{nijkamp2022codegen}
E.~Nijkamp, B.~Pang, H.~Hayashi, L.~Tu, H.~Wang, Y.~Zhou, S.~Savarese, and C.~Xiong.
\newblock Codegen: An open large language model for code with multi-turn program synthesis.
\newblock \emph{arXiv preprint arXiv:2203.13474}, 2022.

\bibitem[Nijkamp et~al.(2023)Nijkamp, Hayashi, Xiong, Savarese, and Zhou]{nijkamp2023codegen2}
E.~Nijkamp, H.~Hayashi, C.~Xiong, S.~Savarese, and Y.~Zhou.
\newblock Codegen2: Lessons for training llms on programming and natural languages, 2023.

\bibitem[OpenAI(2023)]{openai2023gpt4}
OpenAI.
\newblock Gpt-4 technical report, 2023.

\bibitem[Patel et~al.(2021)Patel, Bhattamishra, and Goyal]{patel2021nlp}
A.~Patel, S.~Bhattamishra, and N.~Goyal.
\newblock Are nlp models really able to solve simple math word problems?
\newblock In \emph{Proceedings of the 2021 Conference of the North American Chapter of the Association for Computational Linguistics: Human Language Technologies}, pages 2080--2094, 2021.

\bibitem[Raffel et~al.(2023)Raffel, Shazeer, Roberts, Lee, Narang, Matena, Zhou, Li, and Liu]{raffel2023exploring}
C.~Raffel, N.~Shazeer, A.~Roberts, K.~Lee, S.~Narang, M.~Matena, Y.~Zhou, W.~Li, and P.~J. Liu.
\newblock Exploring the limits of transfer learning with a unified text-to-text transformer, 2023.

\bibitem[Rajbhandari et~al.(2020)Rajbhandari, Rasley, Ruwase, and He]{rajbhandari2020zero}
S.~Rajbhandari, J.~Rasley, O.~Ruwase, and Y.~He.
\newblock Zero: Memory optimizations toward training trillion parameter models.
\newblock In \emph{SC20: International Conference for High Performance Computing, Networking, Storage and Analysis}, pages 1--16. IEEE, 2020.

\bibitem[Roziere et~al.(2023)Roziere, Gehring, Gloeckle, Sootla, Gat, Tan, Adi, Liu, Remez, Rapin, et~al.]{roziere2023code}
B.~Roziere, J.~Gehring, F.~Gloeckle, S.~Sootla, I.~Gat, X.~E. Tan, Y.~Adi, J.~Liu, T.~Remez, J.~Rapin, et~al.
\newblock Code llama: Open foundation models for code.
\newblock \emph{arXiv preprint arXiv:2308.12950}, 2023.

\bibitem[Sakaguchi et~al.(2021)Sakaguchi, Bras, Bhagavatula, and Choi]{sakaguchi2021winogrande}
K.~Sakaguchi, R.~L. Bras, C.~Bhagavatula, and Y.~Choi.
\newblock Winogrande: An adversarial winograd schema challenge at scale.
\newblock \emph{Communications of the ACM}, 64\penalty0 (9):\penalty0 99--106, 2021.

\bibitem[Sennrich et~al.(2015)Sennrich, Haddow, and Birch]{sennrich2015neural}
R.~Sennrich, B.~Haddow, and A.~Birch.
\newblock Neural machine translation of rare words with subword units.
\newblock \emph{arXiv preprint arXiv:1508.07909}, 2015.

\bibitem[Su et~al.(2023)Su, Lu, Pan, Murtadha, Wen, and Liu]{su2023roformer}
J.~Su, Y.~Lu, S.~Pan, A.~Murtadha, B.~Wen, and Y.~Liu.
\newblock Roformer: Enhanced transformer with rotary position embedding, 2023.

\bibitem[Suzgun et~al.(2022)Suzgun, Scales, Sch{\"a}rli, Gehrmann, Tay, Chung, Chowdhery, Le, Chi, Zhou, , and Wei]{suzgun2022challenging}
M.~Suzgun, N.~Scales, N.~Sch{\"a}rli, S.~Gehrmann, Y.~Tay, H.~W. Chung, A.~Chowdhery, Q.~V. Le, E.~H. Chi, D.~Zhou, , and J.~Wei.
\newblock Challenging big-bench tasks and whether chain-of-thought can solve them.
\newblock \emph{arXiv preprint arXiv:2210.09261}, 2022.

\bibitem[Taori et~al.(2023)Taori, Gulrajani, Zhang, Dubois, Li, Guestrin, Liang, and Hashimoto]{alpaca}
R.~Taori, I.~Gulrajani, T.~Zhang, Y.~Dubois, X.~Li, C.~Guestrin, P.~Liang, and T.~B. Hashimoto.
\newblock Stanford alpaca: An instruction-following llama model.
\newblock \url{https://github.com/tatsu-lab/stanford_alpaca}, 2023.

\bibitem[Touvron et~al.(2023)Touvron, Martin, Stone, Albert, Almahairi, Babaei, Bashlykov, Batra, Bhargava, Bhosale, et~al.]{touvron2023llama}
H.~Touvron, L.~Martin, K.~Stone, P.~Albert, A.~Almahairi, Y.~Babaei, N.~Bashlykov, S.~Batra, P.~Bhargava, S.~Bhosale, et~al.
\newblock Llama 2: Open foundation and fine-tuned chat models.
\newblock \emph{arXiv preprint arXiv:2307.09288}, 2023.

\bibitem[Wang et~al.(2021)Wang, Wang, Joty, and Hoi]{wang2021codet5}
Y.~Wang, W.~Wang, S.~Joty, and S.~C. Hoi.
\newblock Codet5: Identifier-aware unified pre-trained encoder-decoder models for code understanding and generation.
\newblock \emph{arXiv preprint arXiv:2109.00859}, 2021.

\bibitem[Zellers et~al.(2019)Zellers, Holtzman, Bisk, Farhadi, and Choi]{zellers2019hellaswag}
R.~Zellers, A.~Holtzman, Y.~Bisk, A.~Farhadi, and Y.~Choi.
\newblock Hellaswag: Can a machine really finish your sentence?
\newblock In \emph{Proceedings of the 57th Annual Meeting of the Association for Computational Linguistics}, pages 4791--4800, 2019.

\bibitem[Zheng et~al.(2023)Zheng, Xia, Zou, Dong, Wang, Xue, Shen, Wang, Wang, Li, et~al.]{zheng2023codegeex}
Q.~Zheng, X.~Xia, X.~Zou, Y.~Dong, S.~Wang, Y.~Xue, L.~Shen, Z.~Wang, A.~Wang, Y.~Li, et~al.
\newblock Codegeex: A pre-trained model for code generation with multilingual benchmarking on humaneval-x.
\newblock In \emph{Proceedings of the 29th ACM SIGKDD Conference on Knowledge Discovery and Data Mining}, pages 5673--5684, 2023.

\end{thebibliography}
